\documentclass[prd,eqsecnum,preprint,tightenlines,nofootinbib]{revtex4}
\usepackage{graphicx,amsmath}

\def\be{\begin{equation}}
\def\bea{\begin{eqnarray}}
\def\ee{\end{equation}}
\def\eea{\end{eqnarray}}

\begin{document}

\preprint{UMN-D-03-3}
\preprint{hep-th/0304147}

\title{Spectrum of ${\cal N}=1$ massive super Yang--Mills theory
with fundamental matter in 1+1 dimensions}

\author{J.R.~Hiller}
\affiliation{Department of Physics,
University of Minnesota-Duluth,
Duluth, MN  55812, USA}

\author{S.S.~Pinsky}
\author{U.~Trittmann}
\affiliation{Department of Physics,
Ohio State University,
Columbus, OH 43210, USA}

\date{\today}

%----------------------------------------------------------------------%
%                       A B S T R A C T
%----------------------------------------------------------------------%

\begin{abstract}
We consider ${\cal N}=1$ supersymmetric Yang--Mills theory with
fundamental matter in the large-$N_c$ approximation in 1+1 dimensions.
We add a Chern--Simons term to give the adjoint partons a mass and solve
for the meson bound states. Here ``mesons" are color-singlet states with two
partons in the fundamental representation. The spectrum is exactly
supersymmetric, and there is complete degeneracy between the fermion and
boson bound states. We find that the mass spectrum is composed of two
distinct bands. We analyze the properties of the bound states in each
band and find a number of interesting properties of these states. In both
bands, some of the states  are nearly pure quark-gluon bound
states while others are nearly pure squark-gluon bound states. The structure
functions of many of the bound states found are very strongly peaked near
$x=0$. The convergence of the numerical approximation appears to be very
good in all cases.
\end{abstract}


\maketitle

%%%%%%%%%%%%%%%%%%%%%%%%%%%%%%%%%%%%%%%%%%%%%
\section{Introduction}
%%%%%%%%%%%%%%%%%%%%%%%%%%%%%%%%%%%%%%%%%%%%%

At this point there is no direct evidence for supersymmetry. Nevertheless
supersymmetry is such a beautiful symmetry and  provides such elegant
solutions to a host of theoretical problems~\cite{TPISUSY} that many believe
that it must be present in nature. It is a pressing experimental issue to
see if nature takes advantage of this elegant option. Of course, we already 
know that supersymmetry is rather badly broken, since we do not see any
superpartners for the particles of the standard model.  It is assumed that
all of the superpartners are in fact very heavy and that we will see them as
we go to higher energies in accelerators. It is therefore extremely
interesting to investigate what the properties of supersymmetric bound states 
might look like. There are indications that some unusual things can
happen in a theory with supersymmetry~\cite{Hiller:2002cu,Hiller:2002pj}.

We present here the solution of an ${\cal N}=1$ super Yang--Mills (SYM)
theory with fundamental matter in 1+1 dimensions. We also add a 
Chern--Simons (CS) term to give the adjoint matter a mass. There are 
several motivations to look at (1+1)-dimensional models. 
't Hooft~\cite{thooft} showed long ago that two-dimensional models 
can be powerful laboratories for the study of the
bound-state problem. These models remain popular to this day because they
are easy to solve and share many properties with their four-dimensional
cousins, most notably stable bound states.  Supersymmetric two-dimensional
models are particularly attractive, since they are also
super-renormalizable.  Given that the dynamics of the gauge field is 
responsible for the strong interaction and for the formation of bound 
states, it comes as no surprise that a great deal of effort has gone 
into the investigation of bound states of pure glue in supersymmetric 
models~\cite{lpreview,alp2}.  Extensive
study of the meson spectrum of non-supersymmetric theories has been done
(see~\cite{bpp98} for a review).  Recently an initial study addressed some
of these states in the context of supersymmetric models~\cite{Lunin:2001im}.
In addition to these reasons for studying (1+1)-dimensional models, we are
also interested in investigating them as the first step toward solving
(2+1)-dimensional models.  A (2+1)-dimensional model can be much more
realistic, since it includes some transverse dynamics.

Currently there  are several numerical approaches to solving field theories.
For QCD-like  theories, lattice gauge theory is probably the most popular
approach, since the lattice approximation does not break the most important
symmetry, gauge symmetry.  Similarly for supersymmetric
theories, supersymmetric discrete light-cone
quantization (SDLCQ)~\cite{bpp98,sakai,lpreview,alp98a} is probably the
most powerful approach since the discretization does not break the most
important symmetry, supersymmetry~\cite{LatticeSUSY}. In this paper we
consider supersymmetric theories and follow this latter approach. To
simplify the calculation we will consider only the large-$N_c$ 
limit~\cite{thooft}, which has proven to be a powerful approximation 
for bound-state calculations.
While baryons can be constructed in this limit~\cite{WittBar}, they have an
infinite number of partons, and thus practical calculations for such states
are complicated.

Throughout this paper we use the word ``meson''
to indicate the group structure of the state.  Namely, we define a meson 
as a bound state whose wave function can be written as a linear combination 
of parton chains, each chain starting and ending with a creation operator 
in the fundamental representation.  In supersymmetric theories, the states 
defined this way can have either bosonic or fermionic statistics.

Previously we saw that the lightest bound states in ${\cal N}=1$
supersymmetric theories are very
interesting~\cite{Hiller:2002cu,Hiller:2002pj}.
In SYM theories in 1+1 and 2+1 dimensions,
the lightest bound states in the spectrum are massless
Bogomol'nyi--Prasad--Sommerfield (BPS) bound states~\cite{lpreview}. These
states are exactly massless at all couplings.  When we add a CS
term to the (1+1)-dimensional SYM theory, which gives a mass to the
constituents, we find approximate BPS states. The masses of these states
are approximately independent of the coupling, and at strong coupling these
states are the lightest bound states in the theory~\cite{Hiller:2002cu}.
In (2+1)-dimensional SYM-CS theory we also find that at strong
coupling there are anomalously light bound states~\cite{Hiller:2002pj}. In
both 2+1 and 1+1 dimensions, these interesting states appear because of the
exact BPS symmetry in the underlying SYM theory.
We have also looked at the lightest bound states of SYM-CS theory  with
fundamental matter~\cite{Hiller:2003qe}. Again we see that the lightest
bound states are significantly lighter than one would have naively expected. 
We found that the lightest state is nearly massless and well below threshold.

We will see that this model has a number of interesting bound states in
addition to the lowest mass state that we studied
previously~\cite{Hiller:2003qe}.  The bound states separate into two bands,
a low-mass band and a high-mass band.  Interestingly the low mass permits
two solutions for bound states. The preferred
solution has an unusual oscillatory convergence.  Some of the states in the
low-mass band are very light and well below threshold. The upper band has
the standard linear convergence.  In both bands, some of the states are nearly
pure quark-gluon states, and some are nearly pure squark and gluons. Some have
structure functions that are very sharply peaked at small longitudinal
momentum fractions, and some have several of these properties.

Throughout this paper we completely ignore the zero-mode
problem~\cite{zm5,thorn}; however, it is clear that considerable progress
on this issue could be made following our earlier work on the zero modes
of the two-dimensional supersymmetric model with only adjoint
fields~\cite{alptzm}.

The paper has the following organization. In Sec.~\ref{SectDefin} we
consider three-dimensional supersymmetric QCD (SQCD) with a CS term and 
dimensionally reduce it to 1+1 dimensions.  We perform the light-cone 
quantization of the resulting  theory by applying canonical commutation 
relations at fixed $x^+\equiv(x^0+x^1)/\sqrt{2}$ and choosing the 
light-cone gauge ($A^+=0$) for the vector field.  After solving the  
constraint equations, we obtain a model containing four dynamical fields.  
We construct the supercharge for this dimensionally reduced theory.  
In Sec.~\ref{SecMesons} we discuss the structure of the lighter 
meson bound states in the large-$N_c$ approximation.
In Sec.~\ref{sec:SuperCS} we discuss the addition of a CS term to the
supercharge~\cite{CSSYM1+1,dunne,witten2}. We explain that, in the context
of this model, this is equivalent to adding a mass to the partons in the
adjoint representation. In Sec.~\ref{secnum} we discuss the bound-state
solutions that we find, including the mass spectrum, structure functions and
the numerical convergence of the various bound states. In Sec.~\ref{secdis}
we discuss our results and the future directions that are indicated by this
research.

%%%%%%%%%%%%%%%%%%%%%%%%%%%%%%%%%%%%%%%%%%%%%%%%%%%%%%%%%
\section{Supersymmetric systems with fundamental matter}
\subsection{Construction of the supercharge}
\label{SectDefin}
%%%%%%%%%%%%%%%%%%%%%%%%%%%%%%%%%%%%%%%%%%%%%%%%%%%%%%%%%

We consider the supersymmetric models in two dimensions which
can  be obtained as the result of dimensional reduction of SQCD$_{2+1}$.
Our starting point is the three-dimensional action
\bea\label{action}
S&=&\int d^3x\mbox{tr}\left(-\frac{1}{4}F_{\mu\nu}F^{\mu\nu}+
\frac{i}{2}{\bar\Lambda}
\Gamma^\mu D_\mu \Lambda +D_\mu \xi^\dagger D^\mu \xi+
i{\bar\Psi} D_\mu\Gamma^\mu\Psi\right.\nonumber\\
&&-\left.g\left[{\bar\Psi}\Lambda\xi+
\xi^\dagger{\bar\Lambda}\Psi\right]\right)\,.
\eea
This action describes a system of a gauge field $A_\mu$, representing
gluons, and its superpartner $\Lambda$, representing gluinos, both taking
values in the adjoint representation of $SU(N_c)$,
and two complex fields, a scalar $\xi$ representing squarks and a Dirac
fermion $\Psi$ representing quarks, transforming according to the
fundamental representation of the same group. In matrix notation the 
covariant derivatives are given by
\be
D_\mu\Lambda=\partial_\mu\Lambda+ig[A_\mu,\Lambda]\,,\quad
D_\mu\xi=\partial_\mu\xi+igA_\mu\xi\,,\quad
D_\mu\Psi=\partial_\mu\Psi+igA_\mu\Psi\,.
\ee
The action (\ref{action}) is invariant under the following supersymmetry
transformations, which are parameterized by a two-component Majorana fermion
$\varepsilon$:
\bea
&&\delta A_\mu=\frac{i}{2}{\bar\varepsilon}\Gamma_\mu\Lambda\,,\qquad
\delta\Lambda=\frac{1}{4}F_{\mu\nu}\Gamma^{\mu\nu}\varepsilon\,,\nonumber\\
&&\delta\xi=\frac{i}{2}{\bar\varepsilon}\Psi\,,\qquad
\delta\Psi=-\frac{1}{2}\Gamma^\mu\varepsilon D_\mu\xi\,.
\eea
Using standard techniques one can construct the Noether current
corresponding to these transformations as
\bea\label{sucurrent}
{\bar\varepsilon}q^\mu&=&\frac{i}{4}{\bar\varepsilon}\Gamma^{\alpha\beta}
\Gamma^\mu\mbox{tr}\left(\Lambda F_{\alpha\beta}\right)+
\frac{i}{2}D^\mu\xi^\dagger\
{\bar\varepsilon}\Psi+\frac{i}{2}\xi^\dagger{\bar\varepsilon}\Gamma^{\mu\nu}
D_\nu\Psi\nonumber\\
&&-\frac{i}{2}{\bar\Psi}\varepsilon D^\mu\xi+\frac{i}{2}D_\nu
{\bar\Psi}\Gamma^{\mu\nu}\varepsilon\xi\,.
\eea

We will consider the reduction of this system to two dimensions, which means
that the field configurations are assumed to be independent of the
transverse coordinate $x^2$. In the resulting two-dimensional system we will
implement light-cone quantization, where the initial conditions as
well as canonical commutation relations are imposed on a light-like
surface $x^+=\mbox{\em const}$. In particular, we construct the supercharge
by integrating the current (\ref{sucurrent}) over the light-like surface to
obtain
\bea\label{sucharge}
{\bar\varepsilon}Q&=&\int dx^-dx^2\left(
\frac{i}{4}{\bar\varepsilon}\Gamma^{\alpha\beta}
\Gamma^+\mbox{tr}\left(\Lambda F_{\alpha\beta}\right)+
\frac{i}{2}D_-\xi^\dagger\
{\bar\varepsilon}\Psi+\frac{i}{2}\xi^\dagger{\bar\varepsilon}\Gamma^{+\nu}
D_\nu\Psi\right.\nonumber\\
&&-\left.\frac{i}{2}{\bar\Psi}\varepsilon D^+\xi+\frac{i}{2}D_\nu
{\bar\Psi}\Gamma^{+\nu}\varepsilon\xi\right)\,.
\eea
Since all fields are assumed to be independent of $x^2$, the integration
over this coordinate gives just a constant factor, which we absorb by a
field redefinition.

If we use the following specific representation for the Dirac matrices in
three dimensions:
\be
\Gamma^0=\sigma_2\,,\qquad \Gamma^1=i\sigma_1\,,\qquad \Gamma^2=i\sigma_3\,,
\ee
the Majorana fermion $\Lambda$ can be chosen to be real.  It is also
convenient to write the fermion fields and the supercharge in component
form as
\be
\Lambda=\left(\lambda,{\tilde\lambda}\right)^T\,,\qquad
\Psi=\left(\psi,{\tilde\psi}\right)^T\,,\qquad
Q=\left(Q^+,Q^-\right)^T\,.
\ee
In terms of this decomposition the superalgebra has an explicit $(1,1)$ form
\be\label{sualg}
\{Q^+,Q^+\}=2\sqrt{2}P^+\,,\qquad \{Q^-,Q^-\}=2\sqrt{2}P^-\,,\qquad
\{Q^+,Q^-\}=0\,.
\ee
The SDLCQ method exploits this superalgebra by constructing $P^-$
from a discrete approximation to $Q^-$~\cite{sakai}, rather than
directly discretizing $P^-$, as is done in ordinary DLCQ~\cite{bpp98}.

To begin to eliminate nondynamical fields, we impose the light-cone gauge
($A^+=0$).  Then the supercharges are given by
\bea\label{Qplus}
Q^+&=&2\int dx^-\left(\lambda\partial_-A^2+
\frac{i}{2}\partial_-\xi^\dagger\psi-\frac{i}{2}\psi^\dagger\partial_-\xi-
\frac{i}{2}\xi^\dagger\partial_-\psi+\frac{i}{2}\partial_-\psi^\dagger\xi
\right)\,,\\
Q^-&=&-2\int dx^-\left(-\lambda\partial_-A^-+
i\xi^\dagger D_2\psi-iD_2\psi^\dagger\xi+\frac{i}{\sqrt{2}}
\partial_-({\tilde\psi}^\dagger\xi-\xi^\dagger{\tilde\psi})\right)
\,.\nonumber   \\
\eea
Note that apart from  a total derivative these expressions involve only
left-moving components of the fermions ($\lambda$ and $\psi$). In fact, in
the light-cone formulation only these components are dynamical. To see this we
consider the equations of
motion that follow from the action (\ref{action}), in light-cone gauge.
Three of them serve as constraints rather than as dynamical statements;
they are
\bea
\partial_-{\tilde\lambda}&=&-\frac{ig}{\sqrt{2}}
\left([A^2,\lambda]+i\xi\psi^\dagger-i\psi\xi^\dagger\right),\\
\partial_-{\tilde\psi}&=&-\frac{ig}{\sqrt{2}}A^2\psi+
\frac{g}{\sqrt{2}}\lambda\xi\,, \label{fcurrent}
\eea 
and
\be\label{constraint}
\partial^2_-A^-=gJ\,,
\ee
with
\be \label{current}
J\equiv i[A^2,\partial_-A^2]+
\frac{1}{\sqrt{2}}\{\lambda,\lambda\}-ih\partial_-\xi\xi^\dagger+
i\xi\partial_-\xi^\dagger+\sqrt{2}\psi\psi^\dagger\,.
\ee
Apart from the zero-mode problem~\cite{zm5}, one can invert the last
constraint to write the auxiliary field $A^-$ in terms of physical degrees
of freedom. 

In order to solve the bound-state problem $2P^+P^-|M\rangle=M^2|M\rangle$,
we apply the methods of
SDLCQ. Namely we compactify the two-dimensional theory on a
light-like circle ($-L<x^-<L$), and impose periodic boundary conditions on
all physical fields. This leads to the following mode expansions:
\bea
\label{expandA2}
A^2_{ij}(0,x^-)&=&\frac{1}{\sqrt{4\pi}}\sum_{k=1}^{\infty}\frac{1}{\sqrt{k}}
\left(a_{ij}(k)e^{-ik\pi x^-/L}+a^\dagger_{ji}(k)e^{ik\pi x^-/L}\right)\,,\\
\label{expandLambda}
\lambda_{ij}(0,x^-)&=&\frac{1}{2^{\frac{1}{4}}\sqrt{2L}}\sum_{k=1}^{\infty}
\left(b_{ij}(k)e^{-ik\pi x^-/L}+b^\dagger_{ji}(k)e^{ik\pi x^-/L}\right)\,,\\
\label{expandxi}
\xi_i(0,x^-)&=&\frac{1}{\sqrt{4\pi}}\sum_{k=1}^{\infty}\frac{1}{\sqrt{k}}
\left(c_i(k)e^{-ik\pi x^-/L}+{\tilde c}^\dagger_{i}(k)e^{ik\pi
x^-/L}\right)\,,\\
\label{expandPsi}
\psi_{i}(0,x^-)&=&\frac{1}{2^{\frac{1}{4}}\sqrt{2L}}\sum_{k=1}^{\infty}
\left(d_{i}(k)e^{-ik\pi x^-/L}+{\tilde d}^\dagger_{i}(k)e^{ik\pi
x^-/L}\right)\,.
\eea 
We drop the zero modes of the fields; including them
could lead to new and interesting effects (see~\cite{alptzm}, for example),
but this is beyond the scope of this work.

In the light-cone formalism one treats $x^+$ as the time direction, thus
the commutation relations between fields and their momenta are imposed on
the surface $x^+=0$.  For the system under consideration this means that
\bea\label{CanComRelField}
\left[A_{ij}^2(0,x^-),\partial_-A_{kl}^2(0,y^-)\right]&=&
i\left(\delta_{il}\delta_{kj}-\frac{1}{N}\delta_{ij}\delta_{kl}\right)
\delta(x^--y^-)\,,\\
\left\{\lambda_{ij}(0,x^-),\lambda_{kl}(0,y^-)\right\}&=&\sqrt{2}
\left(\delta_{il}\delta_{kj}-\frac{1}{N}\delta_{ij}\delta_{kl}\right)
\delta(x^--y^-)\,,\\
\left[\xi_i(0,x^-),\partial_-\xi_j(0,y^-)\right]&=&
i\delta_{ij}\delta(x^--y^-)\,,\\
\left\{\psi_{i}(0,x^-),\psi_{j}(0,y^-)\right\}&=&\sqrt{2}
\delta_{ij}\delta(x^--y^-)\,.
\eea
These relations can be rewritten in terms of creation and annihilation
operators as
\be
\left[a_{ij},a^\dagger_{kl}\right]=
\left(\delta_{il}\delta_{kj}-\frac{1}{N}\delta_{ij}\delta_{kl}\right)\,,
\quad
\left\{b_{ij},b^\dagger_{kl}\right\}=
\left(\delta_{il}\delta_{kj}-\frac{1}{N}\delta_{ij}\delta_{kl}\right)\,,
\ee
\be
\left[c_{i},c^\dagger_{j}\right]=\delta_{ij}\,,\quad
\left[{\tilde c}_{i},{\tilde c}^\dagger_{j}\right]=\delta_{ij}\,,\quad
\left\{d_{i},{d}^\dagger_{j}\right\}=\delta_{ij} \quad
\left\{{\tilde d}_{i},{\tilde d}^\dagger_{j}\right\}=\delta_{ij}\,.
\ee
In this paper we will discuss numerical results obtained
in the large-$N_c$ limit, i.e.\ we neglect $1/N_c$ terms in the above
expressions. Although $1/N_c$ corrections may have interesting
consequences, they are beyond the scope of this work.

Substituting the result into the expression for the supercharge
and omitting the boundary term, we get
\be\label{Qminus}
Q^-= Q^-_s +Q^-_1 +Q^-_2 +Q^-_3\,.
\ee
where $Q^-_s$ is the supercharge of pure adjoint matter~\cite{sakai}.
The three supercharges that govern the behavior of the
fundamental matter in these states are
\bea
Q^-_1&=&-\frac{g}{\sqrt{2}}\int dx^-\left(
i\sqrt{2}\xi\partial_-\xi^\dagger-i\sqrt{2}\partial_-\xi\xi^\dagger\right)
\frac{1}{\partial_-}\lambda\,, \\
Q^-_2&=&-\frac{g}{\sqrt{2}}\int dx^-\left(
2\psi\psi^\dagger\right)
\frac{1}{\partial_-}\lambda\,, \\
Q^-_3&=&-2g\int dx^-\left(\xi^\dagger A^2\psi+\psi^\dagger A^2\xi\right)\,.
\eea
After substituting the expansions (\ref{expandA2}-\ref{expandPsi}),
one gets the mode decomposition of the supercharge,
\begin{equation}
Q^-_\alpha=\frac{i2^{-1/4}g\sqrt{L}}{\pi} \sum_{k_1,k_2,k_3=1}^\infty
q_\alpha^-(k_1,k_2,k_3)\,,
\end{equation}
with
\begin{eqnarray}
\label{Q1}
q^-_1&=&\frac{(k_2+k_3)}{2k_1\sqrt{k_2 k_3}}
[{\tilde c}^\dagger_i(k_2){\tilde c}_{j}(k_3)\tilde b_{ji}(k_1)
-{\tilde c}^\dagger_i(k_2)b_{ij}^\dagger(k_1){\tilde c}_{j}(k_3)
\nonumber \\
&&+b_{ji}^\dagger(k_1) c^\dagger_i(k_2)c_{j}(k_3)
-c^\dagger_i(k_2)b_{ij}(k_1)c_{j}(k_3) ]\delta_{k_3,k_1+k_2}\,,\\
\label{Q2}
q^-_2&=& \frac{1}{k_1}[
{\tilde d}^\dagger_i(k_2)b^\dagger_{ij}(k_1){\tilde d}_j(k_3)+
{\tilde d}^\dagger_j(k_3){\tilde d}_i(k_2)b_{ij}(k_1) +
\nonumber\\ 
&&d^\dagger_i(k_2)b^\dagger_{ij}(k_1) d_j(k_3)+
d^\dagger_j(k_3) d_i(k_2)b_{ij}(k_1) ]\delta_{k_3,k_1+k_2}\,,\\
\label{Q3}
q^-_3&=&\frac{-i}{2\sqrt{k_2 k_3}} [
( d^\dagger_j(k_1){\tilde c}_{i}(k_3)a_{ij}(k_2)+
{\tilde c}^\dagger_{i}(k_3)a^\dagger_{ij}(k_2)d_j(k_1)+
\nonumber\\ 
&& ({\tilde d}^\dagger_{i}(k_1)a_{ij}(k_2)c_j(k_3)+
a_{ij}^\dagger(k_2)c^\dagger_j(k_3){\tilde d}_{i}(k_1))\delta_{k_1,k_3+k_2}
\nonumber\\
&&({\tilde c}^\dagger_j(k_3)d_{i}(k_1)a_{ij}(k_2)+
d^\dagger_{i}(k_1)a^\dagger_{ij}(k_2){\tilde c}_j(k_3)+
\nonumber\\ 
&&( c^\dagger_{i}(k_3)a_{ij}(k_2){\tilde d}_j(k_1)+
a^\dagger_{ij}(k_2){\tilde d}^\dagger_j(k_1)c_{i}(k_3))\delta_{k_3,k_1+k_2}
]\,.
\end{eqnarray}
We have dropped from these expression the terms that connect the states that
are closed loops of adjoint partons to the string-like meson states. These
interactions are of order $1/\sqrt{N_c}$ and can be neglected in the
large-$N_c$ approximation.

%%%%%%%%%%%%%%%%%%%%%%%%%%%%%%%%%%%%%%%%%%%%%%%%%%%%%%%%%%%%%%%%%%%%%%
\subsection{Meson structure}
\label{SecMesons}
%%%%%%%%%%%%%%%%%%%%%%%%%%%%%%%%%%%%%%%%%%%%%%%%%%%%%%%%%%%%%%%%%%%%%%%%

We will consider here only meson-like states. In the large-$N_c$
approximation these are color-singlet states with exactly two partons 
in the fundamental representation.  The boson bound states will have 
either two bosons or two fermions in the fundamental representation.
In general a boson bound state will have a combination of these types of
contributions.  Because this theory and the numerical formalism are exactly
supersymmetric, for each boson bound state there will be a degenerate bound
state that is a fermion. The fermion bound state
will have one fermion in the fundamental
representation and one boson in the fundamental representation. In the
string interpretation of these theories, such states would correspond
to open strings with freely moving endpoints. In the language of QCD, the
model corresponds to a system of interacting gluons and gluinos which bind
dynamical (s)quarks and anti-(s)quarks. In the large-$N_c$ limit we will
have to consider only a single (s)quark--anti-(s)quark pair.  Thus the Fock 
space is constructed from states of the following type:
\be\label{state}
{\bar f}^\dagger_{i_1}(k_1) a^\dagger_{i_1i_2}(k_2)\dots
b^\dagger_{i_ni_{n+1}}(k_{n-1})\dots f^\dagger_{i_p}(k_n)|0\rangle\,.
\ee
Here $\bar{f}_i^\dagger$ and $f_i^\dagger$ each create one of the
fundamental partons, and $|0\rangle$ is the vacuum annihilated by $a_{ij}$,
$b_{ij}$, $c_i$, ${\tilde c}_i$, $d_i$, and ${\tilde d}_i$.

The other color-singlet bound states in this theory are states that are
composed of traces of only adjoint mesons. These can be considered to be
loops.  At finite $N_c$ this theory has interactions that break these loops 
and insert a pair of fundamental partons, making an open-string state. This 
type of interaction can, of course, also form loops from open strings and 
break one open string into two open strings. In principle, a calculation of 
the spectrum of such a finite-$N_c$ theory is within the reach of SDLCQ. The
only significant change is to include states in the basis with more than
one color trace.

%%%%%%%%%%%%%%%%%%%%%%%%%%%%%%%%%%%%%%%%%%%%%%%%%%%%%%%%%%%%%%%
\subsection{Supersymmetric Chern--Simons theory} \label{sec:SuperCS}
%%%%%%%%%%%%%%%%%%%%%%%%%%%%%%%%%%%%%%%%%%%%%%%%%%%%%%%%%%%%%%%%%%

The CS term we use in this calculation is obtained by starting with a
CS term in 2+1 dimensions and reducing it to 1+1 dimensions.  This
term has the effect of adding a mass for the adjoint partons.
In this calculation we are including fundamental matter because we are
interested in QCD-like meson bound states. Without a mass for the adjoint
matter, this theory is known to produce very
long light chains of adjoint partons. In a finite-$N_c$ calculation we
would not have these very long chains because they would break, but
in the large-$N_c$ approximation they do not. While SDLCQ can be used to do
finite-$N_c$ calculations, it is much easier to add a mass to restrict
the number of adjoint partons in our bound states.  We choose the CS
mechanism to give the adjoint partons a mass because we can do this
without breaking the supersymmetry.

The Lagrangian  of this theory is
\begin{equation} \label{Lagrangian}
{\cal L}={\cal L}_{\rm SQCD}+\frac{\kappa}{2}{\cal L}_{\rm CS}\,,
\end{equation}
where ${\cal L}_{\rm SQCD}$ is the SQCD Lagrangian we discussed earlier,
$\kappa$ is the CS coupling, and
\begin{equation}
{\cal L}_{\rm CS}=\epsilon^{\mu\nu\lambda}\left(A_{\mu}
\partial_{\nu}A_{\lambda}+\frac{2i}{3}gA_\mu A_\nu A_\lambda \right)
+2\bar{\Psi}\Psi\,. \label{eq:CSLagrangian}
\end{equation}
A trace of the color matrices is understood.
The constraint equation (\ref{fcurrent}) gains a third term
of the form $-\kappa\lambda/\sqrt{2}$ on the right-hand side,
and the definition of the current in (\ref{current}) now
has an additional term, $\kappa \partial_- A^2$.
The discrete version of the CS part of the supercharge in 1+1
dimensions is
\begin{equation} \label{qcs}
Q^-_{CS}=\left(\frac{2^{-1/4}\sqrt{L}}{\sqrt{\pi}}\right)
\sum_{n}\frac{\kappa}{\sqrt{n}}
\left(a_{ij}^{\dagger}(n)b_{ij}(n)+b_{ij}^{\dagger}(n)a_{ij}(n)\right)\,.
\end{equation}

It is informative to compare $Q_{\rm CS}^-$ with the one term in the
supercharge for \mbox{${\cal N}=1$} SYM in 2+1 dimensions~\cite{hpt2001} 
which has an explicit dependence on the transverse momentum $k_\perp$.  
This $k_\perp$-dependent term has the form
\begin{equation}
Q^-_{\perp}=i\left(\frac{2^{-1/4}\sqrt{L}}{\sqrt{\pi}}\right)
                \sum_{n,n_\perp}\frac{k_\perp}{\sqrt{n}}
                     \left(a_{ij}^{\dagger}(n,n_\perp)b_{ij}(n,n_\perp)
                         -b_{ij}^{\dagger}(n,n_\perp)a_{ij}(n,n_\perp)\right)\,,
\end{equation}
where $k_\perp=2\pi n_\perp/L_\perp$ is the discrete transverse
momentum.  Notice that $k_\perp$ and $\kappa$ enter the supercharge in
very similar ways.  Addition of a CS term has the effect of changing
$k_\perp$ to $k_\perp+i\kappa$.  Because the light-cone energy is of
the form $(k^2_\perp + m^2)/k^+$, $k_\perp$ behaves like a
mass, and therefore $\kappa$ also behaves in many ways like a
mass for the adjoint particles.

The partons in the fundamental representation in this theory will remain
massless. Of course, in a more physical theory the supersymmetry would be
badly broken; the squark would acquire a large mass, and only the quarks
would remain nearly massless.

%%%%%%%%%%%%%%%%%%%%%%%%%%%%%%%%%%%%%%%%%%%%%%%%%%%%%%%%%%%
\section{Numerical results}
\label{secnum}
%%%%%%%%%%%%%%%%%%%%%%%%%%%%%%%%%%%%%%%%%%%%%%%%%%%%%%%%%%

This SYM-CS theory with fundamental matter has two dimensionful parameters
with dimension of a mass squared, the YM coupling squared $g^2 N_c/\pi$
and the CS coupling squared $\kappa^2$.  The latter is also the mass squared
of the partons in the adjoint representation. All of the masses in this
paper will be given in units of $g \sqrt{N_c/\pi}$, which will usually be
suppressed.  Furthermore, we are only considering meson bound states. These 
are states of the form shown in Eq.~(\ref{state}) with two fundamental partons 
linked by partons in the adjoint representation. Since we are working in 
the large-$N_c$ approximation, this class of states is disconnected from 
the other allowed class of pure adjoint matter bound states and multiparticle 
states. This theory also has a $Z_2$ symmetry~\cite{kutasov93} which is very 
useful in labeling the states and reducing the dimension of the Fock basis 
that one has to consider in any one diagonalization step. For this theory 
the $Z_2$ symmetry divides the basis into states with an even or odd number 
of gluons.

%%%%%%%%%%%%%%%%%%%%%%%%%%%%%%%%%%%%%%%%%%%%%%%%%%%%%%%%%%
\subsection{Spectra}
%%%%%%%%%%%%%%%%%%%%%%%%%%%%%%%%%%%%%%%%%%%%%%%%%%%%%%%%%%

\begin{figure}
\begin{tabular}{cc}
\includegraphics[width=7.5cm]{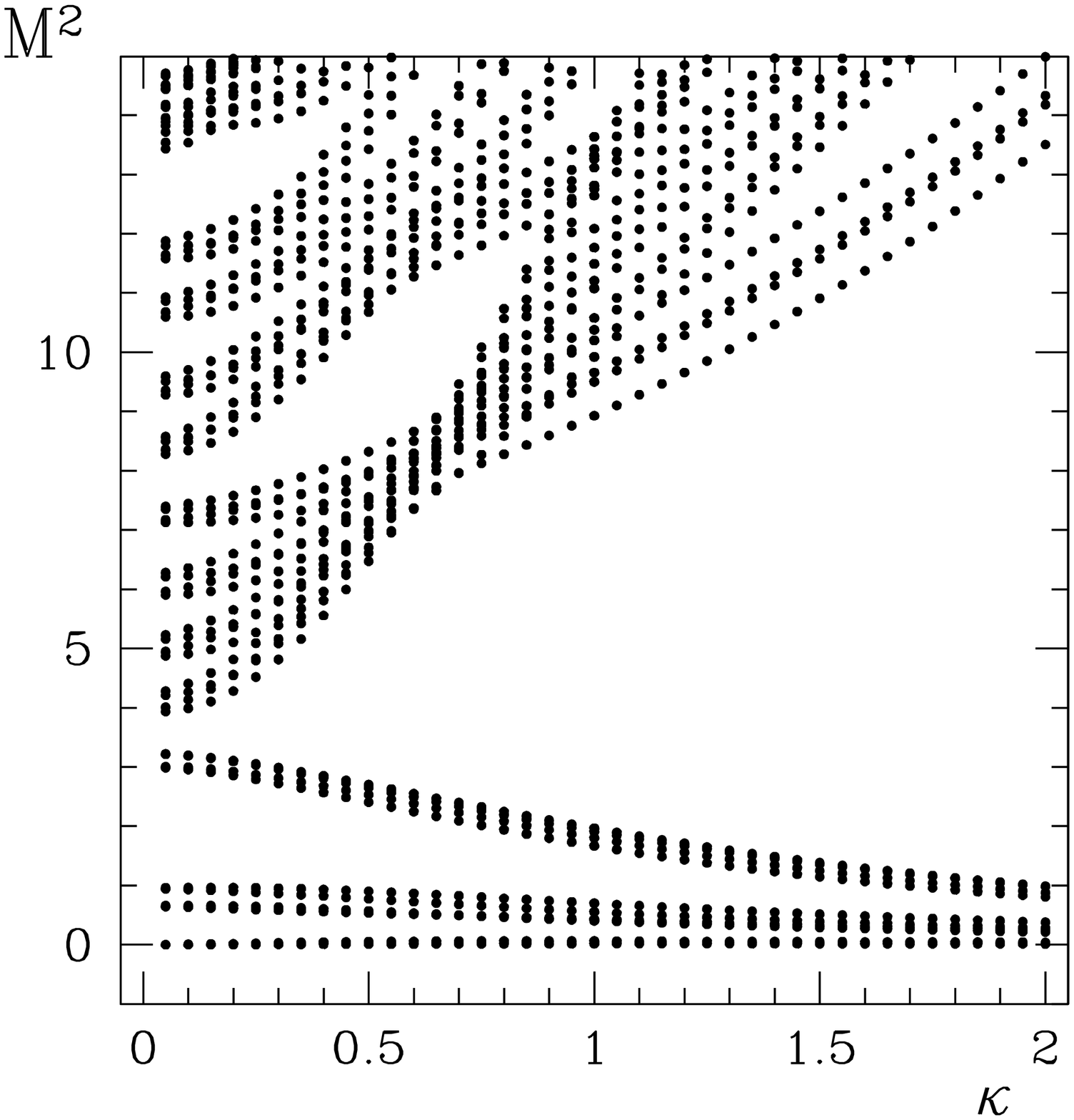} &
\includegraphics[width=7.5cm]{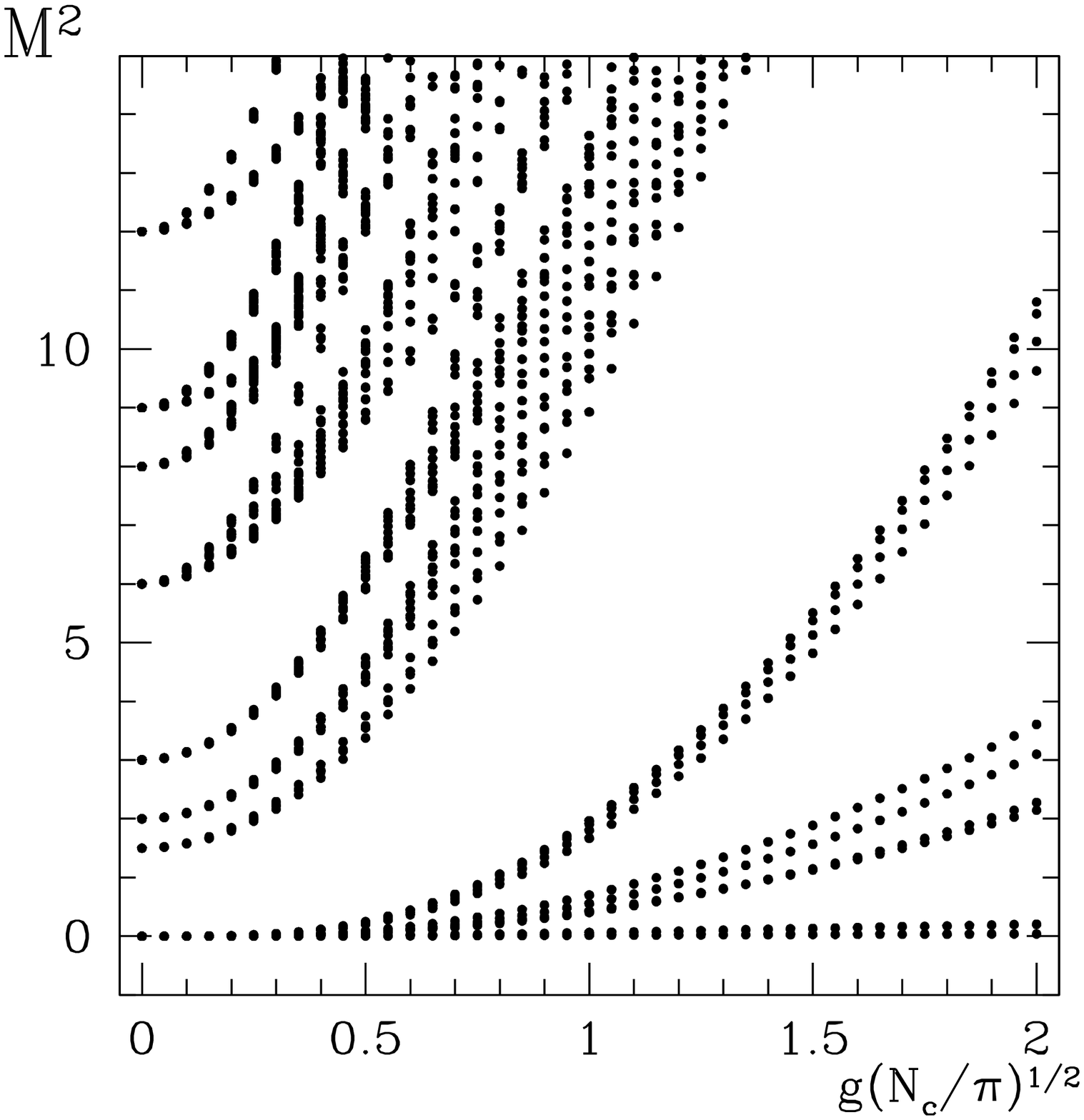} \\
(a) & (b)
\end{tabular}
\caption{The mass-squared spectrum, with $Z_2$ odd, in units of
$g^2N_c/\pi$,
at a resolution of $K=6$ as a function of (a) $\kappa$ at
$g\sqrt{N_c/\pi}=1$
and (b) $g\sqrt{N_c/\pi}$ at $\kappa=1$.}
\label{fig:spectrum}
\end{figure}

We find that the spectra of meson bound states, for both odd and even $Z_2$
symmetry, divides into two bands of states as we increase $\kappa$, a
light-mass band and a heavy-mass band,
as can be seen in Fig.~\ref{fig:spectrum}(a). As $\kappa$ grows, a gap
develops in the spectrum.  This mass gap reflects in part the number of
massive adjoint partons in the bound states.  At very large $\kappa$, the 
low-mass band will comprise states with only fundamental partons. We will 
look at the bound states in both bands and consider various values of 
$\kappa$ relative to $g\sqrt{N_c/\pi}$. In Fig.~\ref{fig:spectrum}(b) we 
see that at $\kappa =1$ this gap grows with the coupling. The lowest mass 
state remains very light even for values of $g\sqrt{N_c/\pi}$ up to 2. As 
far as we can tell, this state remains very light even at very large couplings.
\begin{figure}
\begin{tabular}{cc}
\includegraphics[width=7.5cm]{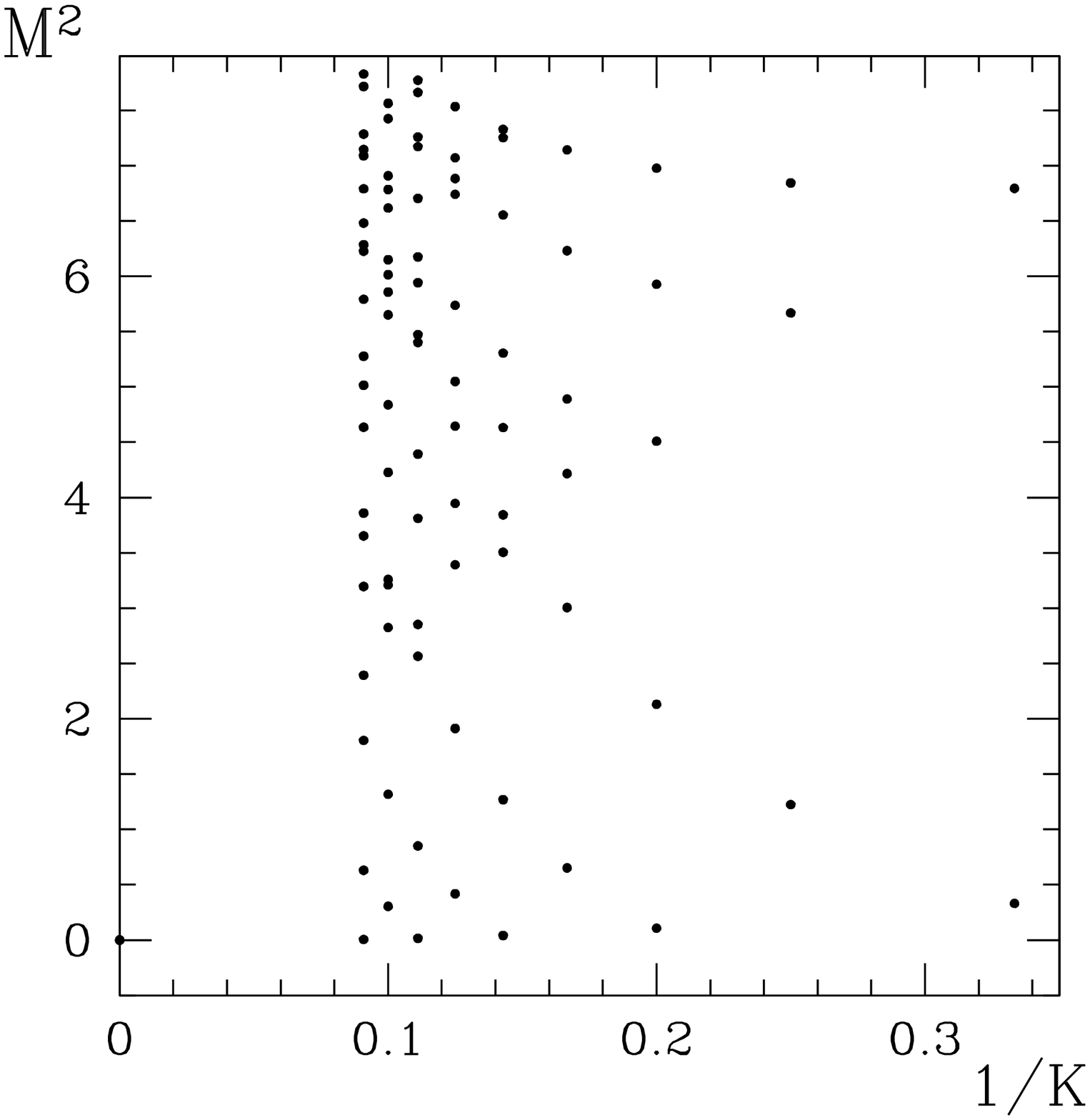}&
\includegraphics[width=7.5cm]{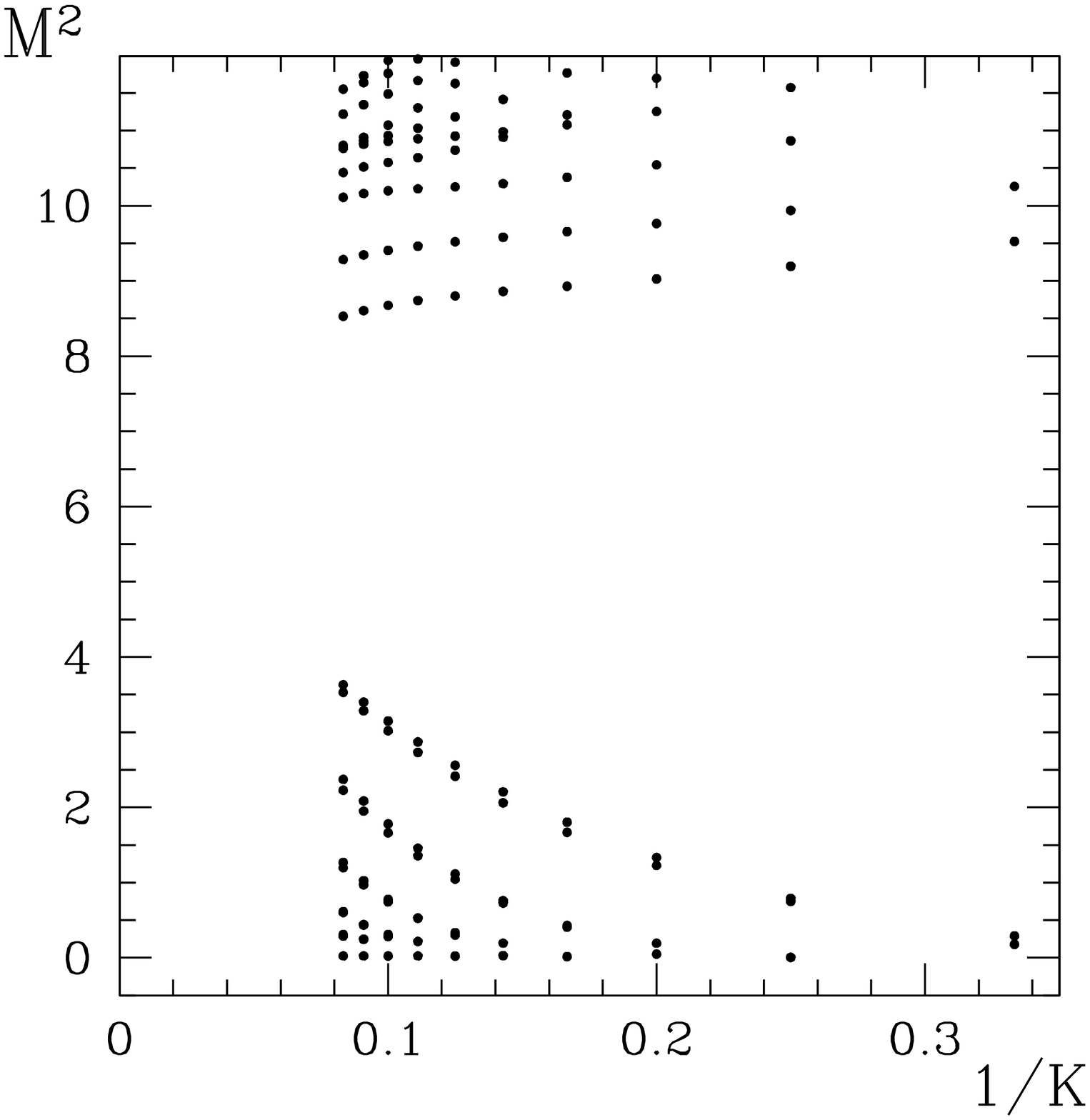}\\
(a) & (b)
\end{tabular}
\caption{The mass-squared spectrum, with $Z_2$ even, in units of
$g^2N_c/\pi$,
as a function of the resolution at (a) $g=0$ and (b) $g\sqrt{N_c/\pi}=1$}.
\label{fig:spectrum2}
\end{figure}

In Fig.~\ref{fig:spectrum2}(a) and (b) we show the spectrum in the
$Z_2$-even sector, with $g\sqrt{N_c/\pi}=1$, as a function of resolution 
at $\kappa=$0 and 1, respectively. We see that at large resolution with 
$\kappa=0$ the two bands merge. It is unclear from this figure whether 
the states in the lower band mix
with the states in the upper band in the region where their masses overlap.
An inspection of adjacent states seems to show that these states remain on
smooth trajectories and do not repel, as one would expect if they were
interacting. This seems to indicate that separate analysis of
these two sets of states is possible. For the most part, however, we will
look at the situation where $\kappa=1$, as shown in
Fig.~\ref{fig:spectrum2}(b). In this case the two bands are well separated
even at the highest accessible resolution, and, therefore, there is not a
problem considering the two bands separately.

Let us first consider the bound states in the lower band. To understand
the physics underlying these bound states we will look at both the spectrum
and the structure functions of individual states. In this theory we have
four species of partons: adjoint fermions and bosons, and fundamental fermions
and bosons. Therefore, we will look at four structure functions that give
the probability of finding a particular variety of parton at a particular
value of longitudinal momentum fraction in a particular bound state. We need
all of this information about each bound state to properly analyze this
theory.

Although the procedure that we follow for identifying bound states is the
standard one in DLCQ, it is worthwhile to give a brief description of that
procedure here because of the unusual nature of the lower band. To identify
a bound state, we start at the lowest value of the resolution $K$, usually
$K=3$, and select a particular eigenvalue, usually the lowest nonzero one.
We then look at the properties of this state, which we always calculate along
with the mass. In principle, we can look at the entire wave function, but for
higher resolution there is more data than we can efficiently handle. It is
efficient to look at the average number of partons of the various types and 
sometimes the average  momentum of some of these partons. We then move to the 
next higher resolution and try to identify an eigenstate that has the same or 
very similar properties to the state at lower resolution. We continue this 
through all the resolutions. This then gives us the mass squared of a 
particular bound state and the various properties of the state as functions 
of the resolution. If this history of a state as a function of the resolution 
makes sense, and can be extrapolated to infinite resolution, we say we have 
found a bound state.  We then repeat the process to obtain another bound 
state. For the oscillatory states presented below, we actually started with 
the lightest state at the highest resolution and then followed the states 
down to low resolution. We then repeated the process with the next highest 
states at the highest resolution.

\begin{figure}
\begin{tabular}{cc}
\includegraphics[width=7.5cm]{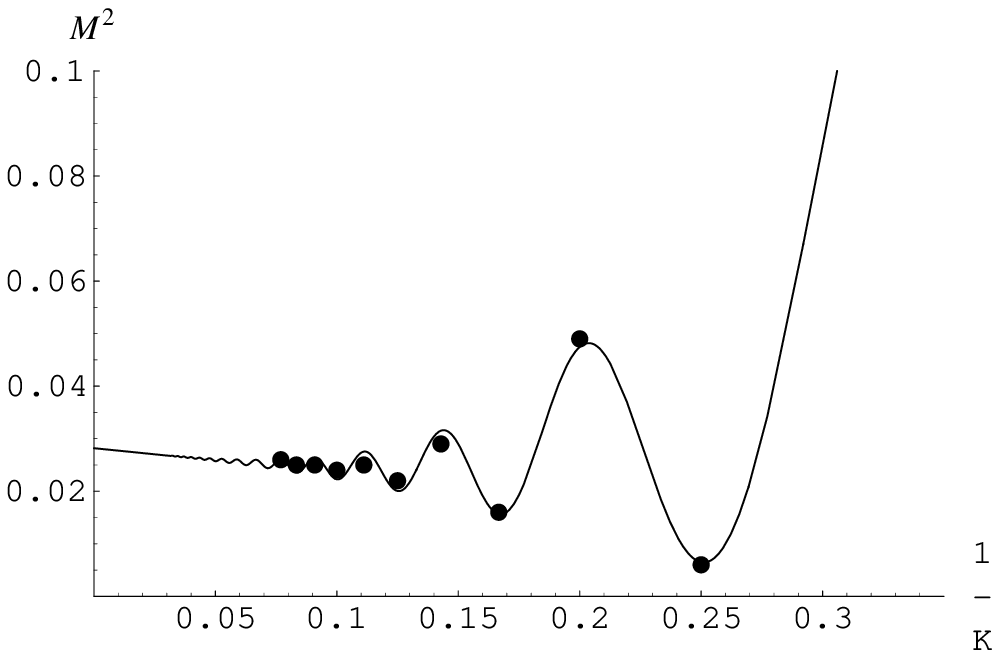} &
\includegraphics[width=7.5cm]{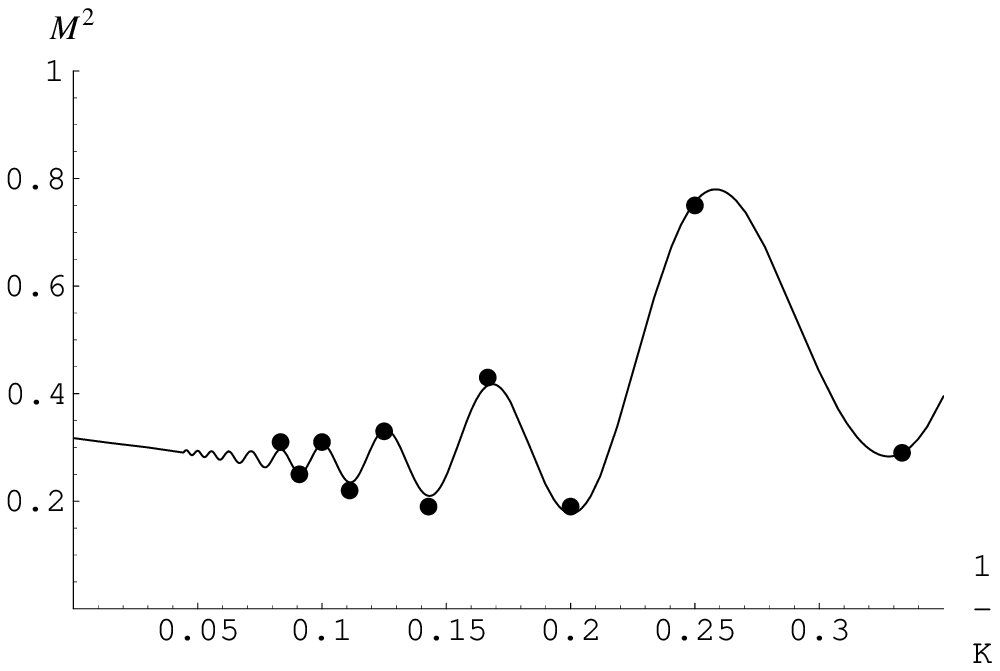} \\
(a) & (b)\\
\includegraphics[width=7.5cm]{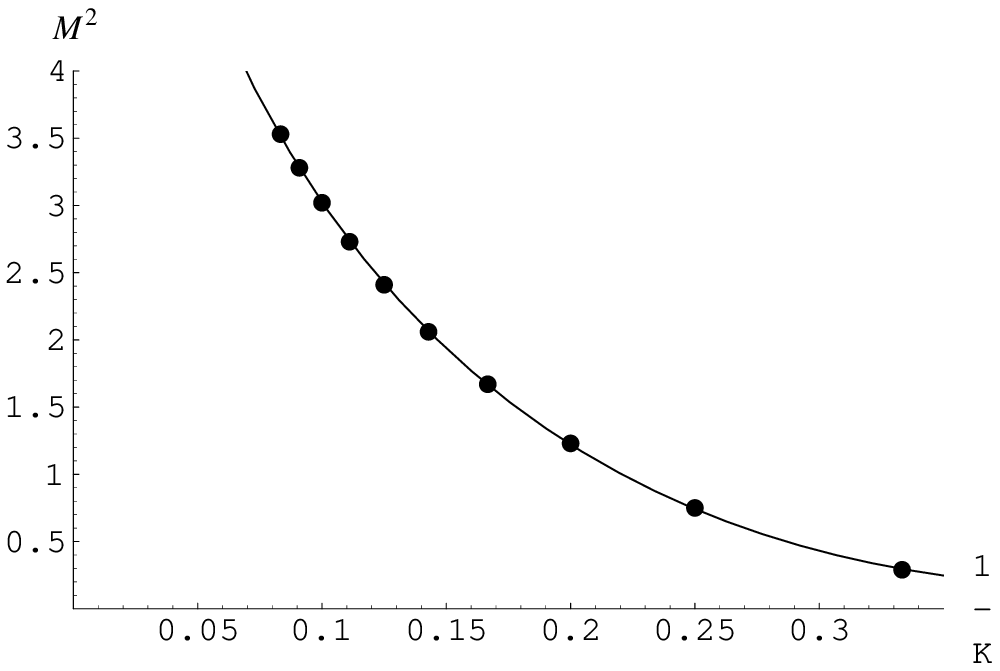}&
\includegraphics[width=7.5cm]{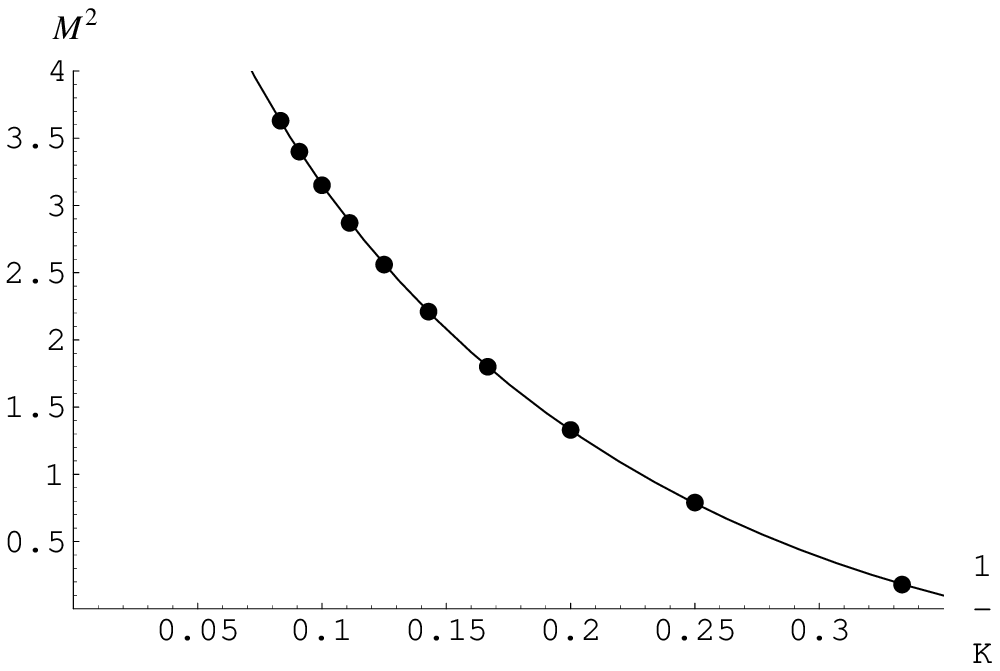}\\
(c) & (d)\\
\end{tabular}
\caption{The mass squared in units of
$g^2N_c/\pi$, as a function of $1/K$
for $\kappa=g\sqrt{N_c/\pi}$ in the $Z_2$-even sector, of
(a) the lowest mass oscillatory state,
(b) the second lowest mass oscillatory state,
(c) the first divergent state, and
(d) the second divergent state.
The solid curve is a fit to the computed points.}
\label{fig:da1}
\end{figure}

The lower band of this model is unique relative to all other models we have
studied in SDLCQ or DLCQ. We find that there are two different ways of
carrying out the above procedure to find the states in the lower band. We
can follow the eigenvalues from resolution to resolution in two ways. As a
result we end up with two sets of bound states, all of whose properties are 
very smoothly behaved as functions of the resolution.  Two such states and 
their properties are shown in Tables~\ref{table1} and \ref{table2}. The 
masses in Table~\ref{table1} are very small.  We have discussed the lowest 
mass state in this spectrum in detail elsewhere~\cite{Hiller:2003qe}. The 
convergence as a function of the resolution is oscillatory, but clearly as 
we move to large resolution the convergence is very good. In 
Fig.~\ref{fig:da1}(a) we present the spectrum as a function of $1/K$, and 
we see that the oscillations lead to
good convergence that extrapolates nicely to $K=\infty$.  These states have
masses near zero. The lowest state has on average one adjoint parton;
therefore, the threshold is at $\kappa^2$.  Thus this is a deeply bound state.  
In Fig.~\ref{fig:da1}(b) we show the spectrum for the next highest mass bound
state, and again we see very good convergence to a very light state. In fact
all of the data in the lower band describe a series of oscillatory states at
higher masses.
\begin{table}
\center{
\begin{tabular}{|c|c|c|c|c|c|c|}
\hline
K & $M^2$ &$<n>$  &$<n_{aB}>$&$<n_{fB}>$&$<n_{aF}>$&$<n_{fF}>$\\
\hline
3    & 0.178    &    2.30    &    0.30    & 1.03    & 0.01    & 0.97\\
4    & 0.006    &    2.56    &    0.51    & 1.86    & 0.05    & 0.14\\
5    & 0.049    &    2.69    &    0.63    & 1.29    & 0.06    & 0.71\\
6    & 0.016    &    2.83    &    0.75    & 1.71    & 0.08    & 0.30\\
7    & 0.029    &    2.84    &    0.76    & 1.45    & 0.08    & 0.55\\
8    & 0.022    &    2.92    &    0.83    & 1.58    & 0.09    & 0.42\\
9    & 0.025    &    2.92    &    0.83    & 1.49    & 0.10    & 0.51\\
10   & 0.024    &    2.96    &    0.86    & 1.52    & 0.10    & 0.48\\
11   & 0.025    &    2.97    &    0.87    & 1.49    & 0.11    & 0.51\\
12   & 0.025    &    3.00    &    0.89    & 1.48    & 0.11    & 0.52\\
13   & 0.026    &    3.01    &    0.90    & 1.47    & 0.11    & 0.53\\
\hline
\end{tabular}
}
\caption{Properties of the lowest mass boson bound state in the $Z_2$-even
sector, including the average numbers of adjoint bosons $aB$, adjoint
fermions $aF$, fundamental bosons $fB$, and fundamental fermions $fF$,
for different values of the resolution $K$.  The mass squared $M^2$ is given
in units of $g^2N_c/\pi$.  The CS coupling is $\kappa=g\sqrt{N_c/\pi}$.}
\label{table1}
\end{table}
\begin{table}
\center{
\begin{tabular}{|c|c|c|c|c|c|c|}
\hline
K & $M^2$ &$<n>$  &$<n_{aB}>$&$<n_{fB}>$&$<n_{aF}>$&$<n_{fF}>$\\
\hline
3   & 0.29  & 2.39  & 0.33  & 1.01  & 0.06  & 0.99  \\
4   & 0.75  & 2.58  & 0.47  & 0.91  & 0.11  & 1.09  \\
5   & 1.23  & 2.73  & 0.59  & 0.87  & 0.14  & 1.14  \\
6   & 1.67  & 2.82  & 0.66  & 0.85  & 0.17  & 1.15  \\
7   & 2.06  & 2.88  & 0.70  & 0.84  & 0.18  & 1.16  \\
8   & 2.41  & 2.92  & 0.72  & 0.84  & 0.20  & 1.16  \\
9   & 2.73  & 2.94  & 0.73  & 0.84  & 0.21  & 1.16  \\
10  & 3.02  & 2.96  & 0.74  & 0.84  & 0.22  & 1.16  \\
11  & 3.28  & 2.97  & 0.74  & 0.84  & 0.23  & 1.16  \\
12  & 3.53  & 2.98  & 0.74  & 0.84  & 0.23  & 1.16  \\
\hline
\end{tabular}
}
\caption{Same as Table~\ref{table1}, but for the lowest mass divergent
state.
%Properties of the lowest mass boson divergent state
%of the $Z_2$-even sector, including the
%average numbers of adjoint bosons $aB$, adjoint fermions $aF$,
%fundamental bosons $fB$, and fundamental fermions $fF$, for different
%values of the resolution $K$.  The mass squared $M^2$ is given in
%units of $g^2N_c/\pi$.  The CS coupling is $\kappa=g\sqrt{N_c/\pi}$.
}
\label{table2}
\end{table}

\begin{figure}
\centerline{
\includegraphics[width=7.5cm]{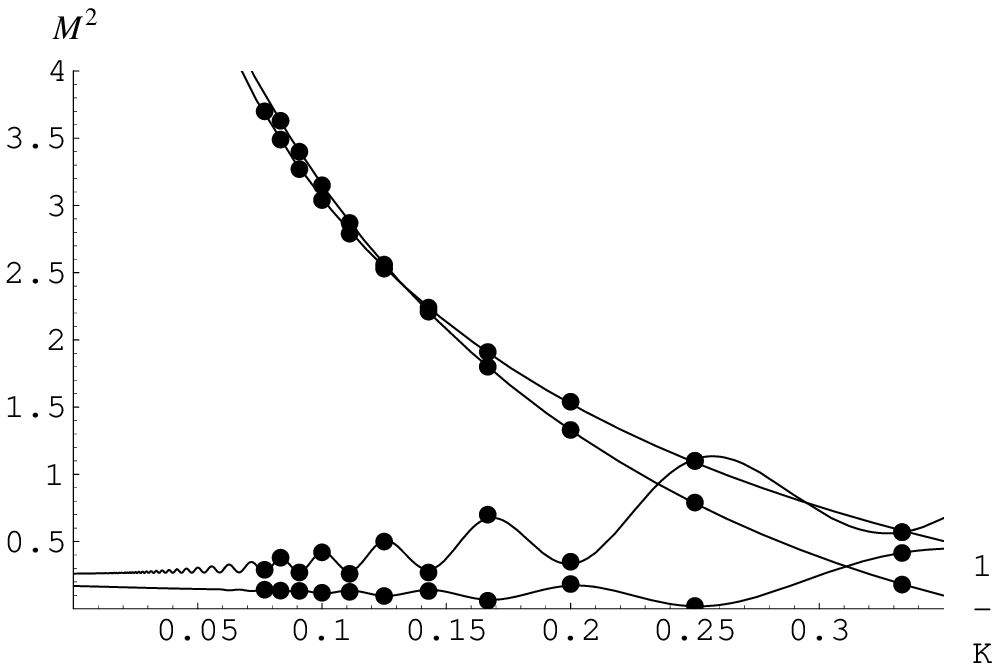}}
\caption{The mass squared in units of
$g^2N_c/\pi$, as a function of $1/K$
for $\kappa=g\sqrt{N_c/\pi}$ in the $Z_2$-even sector of
oscillatory and divergent mass fits.}
\label{fig:da2}
\end{figure}

Shown in Table~\ref{table2} is another way of organizing the same set of
data used to analyze the oscillatory states discussed above, and all these 
states also converge numerically, but to infinite mass. As a function of 
the resolution we see almost no variation in
the properties of this state.  We will refer to these
states as divergent states, and we will argue that they are divergent. This
data and a fit to the data points are shown in Fig.~\ref{fig:da1}(c), and
there are no oscillations.  In Fig.~\ref{fig:da1}(d) we see another state 
of this type. 
It appears that the divergent states do extrapolate to an infinite mass as
$K\rightarrow\infty$. The curves, however, can be fit by a variety of
functions.  In fact the functions $a +b/\sqrt{K} +1/K$ and $a +b\log K +1/K$ 
produce equally good fits. The fits shown are of the former form and have 
intercepts between 10 and 11.  
In Fig.~\ref{fig:da2} we put all four plots from
Fig.~\ref{fig:da1} on the same graph, so that one can
clearly see that at low resolution both curves involve the same data.
In the DLCQ procedure a state must
extrapolate nicely to $K=\infty$, if it is a true bound state of the theory.
It is unclear from the data whether this is true for these states. Our
prejudice is that this bound state diverges as $K \rightarrow \infty$.

\begin{figure}
\begin{tabular}{cc}
\includegraphics[width=7.5cm]{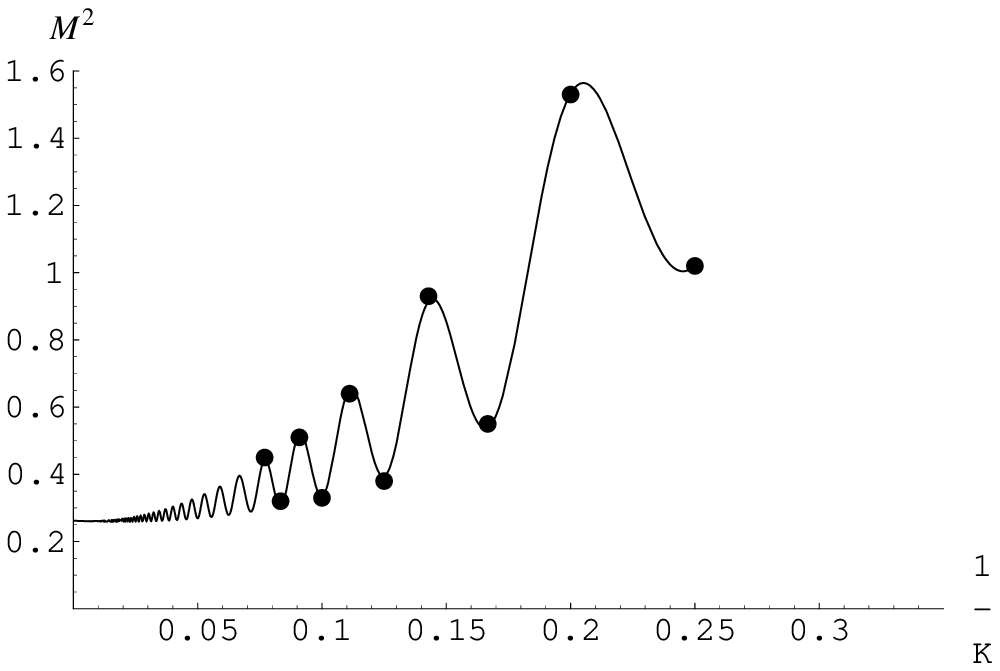}&
\includegraphics[width=7.5cm]{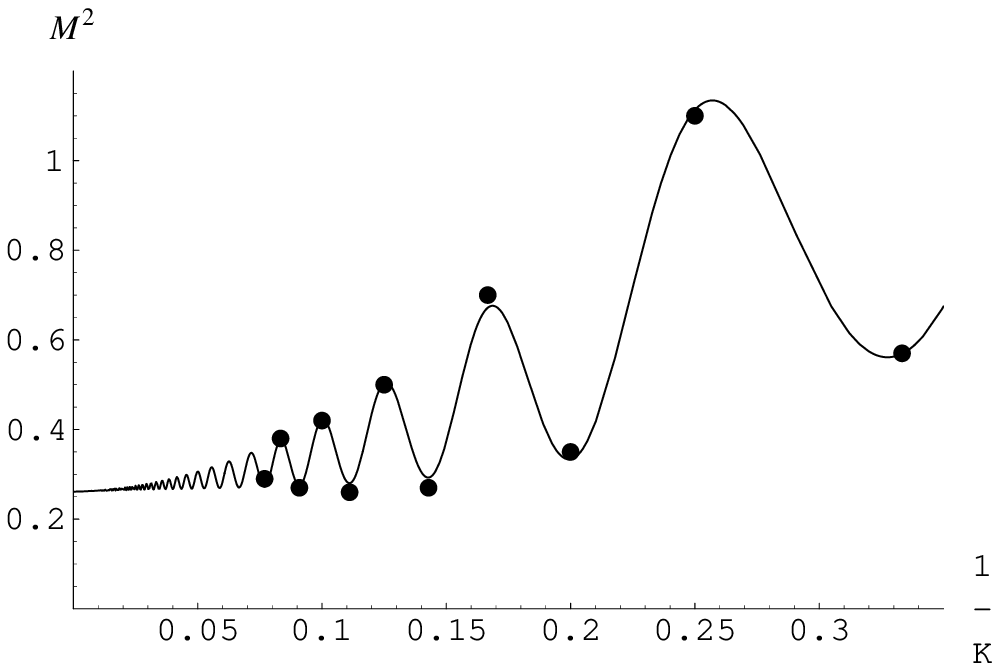}\\
(a) & (b)\\
\includegraphics[width=7.5cm]{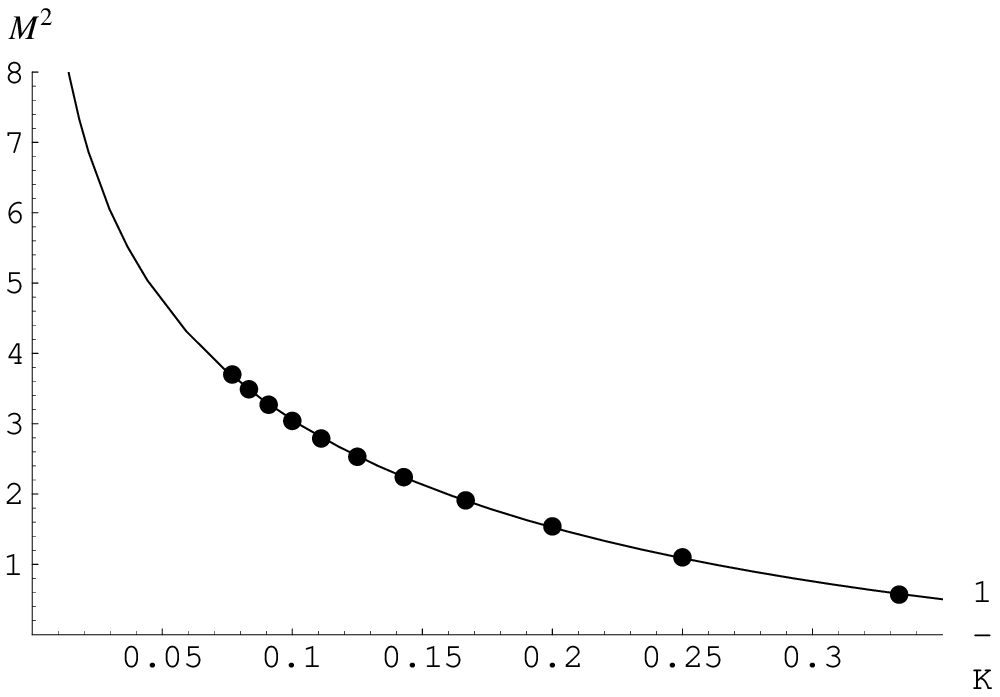}&
\includegraphics[width=7.5cm]{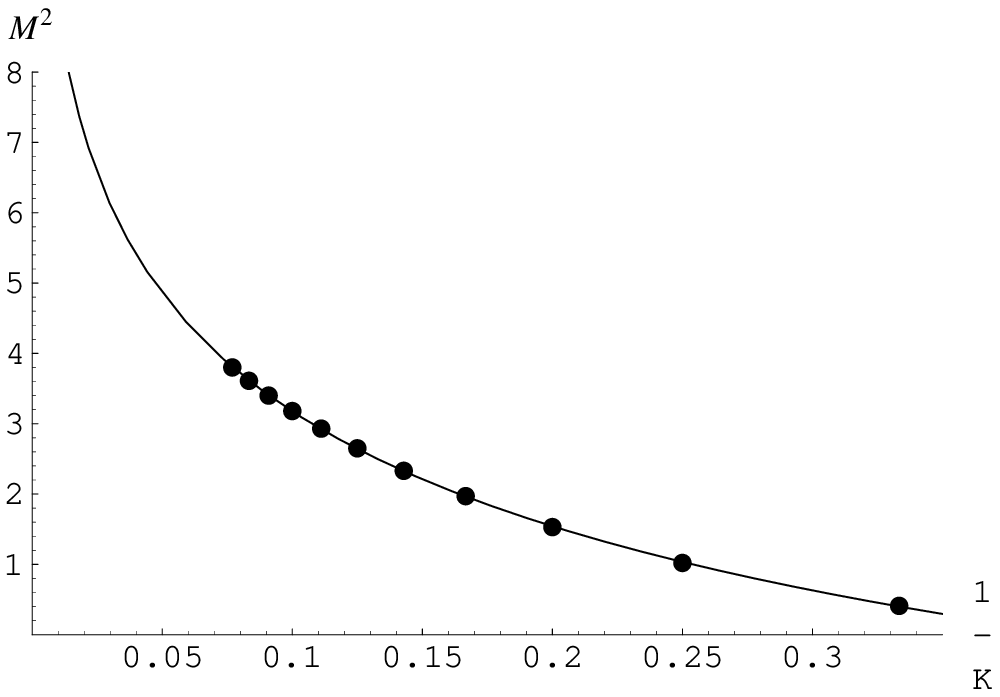}\\
(c) & (d)\\
\end{tabular}
\caption{Same as Fig.~\ref{fig:da1}, but for the $Z_2$-odd sector.
%The mass squared in units of
%$g^2N_c/\pi$, as a function of $1/K$
%for $\kappa=g\sqrt{N_c/\pi}$ in the $Z_2$-odd sector of
%(a) the lowest mass oscillatory state,
%(b) the second lowest mass oscillatory state,
%(c) the first divergent state, and
%(d) the second divergent state.
%The solid curve is a fit to the computed points.
}
\label{fig:da3}
\end{figure}

\begin{figure}
\centerline{
\includegraphics[width=7.5cm]{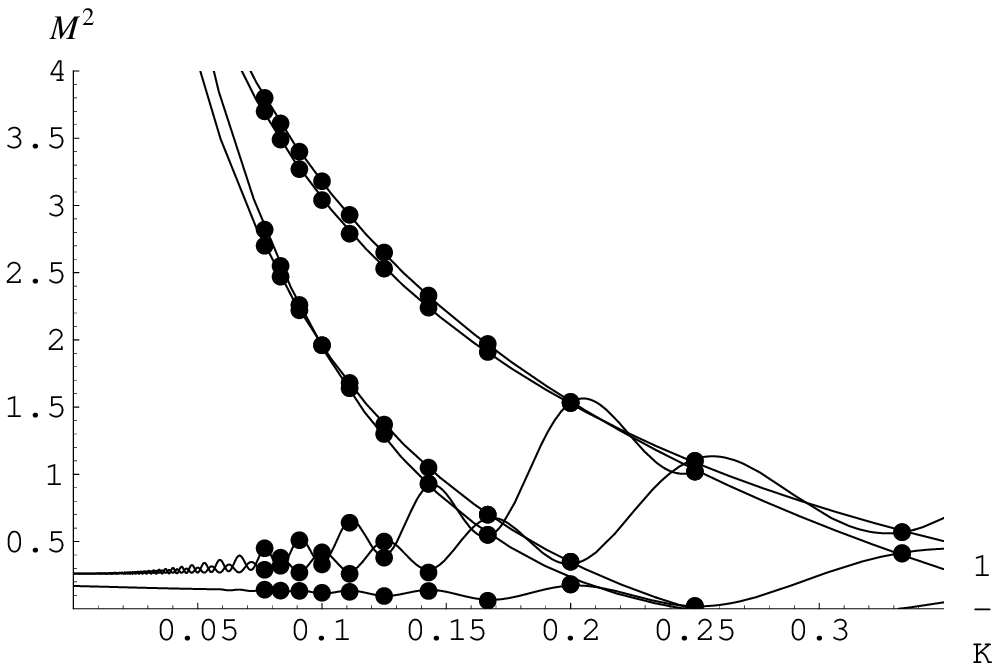}}
\caption{Same as Fig.~\ref{fig:da2}, but for the $Z_2$-odd sector.
%The mass squared in units of
%$g^2N_c/\pi$, as a function of $1/K$
%for $\kappa=g\sqrt{N_c/\pi}$ in the $Z_2$-odd sector of
%oscillatory and divergent mass fits
}
\label{fig:da4}
\end{figure}

In Fig.~\ref{fig:da3} we show similar results for the $Z_2$-odd sector.
In Fig.~\ref{fig:da4} we again show the oscillatory and divergent fits on
the same figure, but now we add a few more states to better show the 
interlacing of these fits to the same data.  There are, of course, 
degenerate fermion bound states for all these states, whose properties 
we do not show.

The oscillatory behavior made this
calculation particularly challenging numerically. We were forced to go to
very high resolution, $K=13$, to be certain that the spectrum really
converged. This was made even more difficult because we had four species 
of particles in the problem.

To get some insight into the behavior of the low-mass band, we recall
the behavior of the two-particle continuum in DLCQ. The ``mass squared''
of two free partons of mass $m$ at resolution $K$ is
\begin{equation}
M^2=m^2 K \left[ \frac{1}{K-n} +\frac{1}{n} \right]\,,
\end{equation}
where one free parton has longitudinal momentum $K-n$ and the other has
$n$~\cite{Gross:1997mx}. In DLCQ this formula produces a band of states.
It is interesting to compare the top and bottom of this band with the top
and bottom of the low-mass band of bound states we have been studying. The
top of the band is obtained by fixing $n$ at one. If we then take $K$ large,
we find $M^2=m^2(K +1 + 1/K +...)$.  Not surprisingly, the divergent curves
in Figs.~\ref{fig:da1}(c) and (d), and Figs.~\ref{fig:da3}(c) and (d) can be
fit with a function of the form $M^2=a +bK+c/K$. In essence, the form of
the divergent state is comparable to the shape of the top of the continuum
band.  Similarly the bottom of the continuum band is obtained by taking 
$n=K/2$ for $K$ even and $n=(K-1)/2$ for $K$ odd. The bottom of the 
continuum band oscillates as a function of $K$. For $K$ even it is $4 m^2$, 
and for $K$ odd it is $4 m^2(1+1/K+...)$. It again appears that the lowest 
mass states resemble the bottom of the continuum band. We would expect to 
see a connection between the weak-coupling bound states and the free 
spectrum, but this connection appears to extend to $g^2 N_c/\pi =\kappa^2$. 
Based on this argument, the divergent states actually do diverge and are 
not true states of the theory.

The lowest mass $Z_2$-even state in the low-mass band has on average one
adjoint parton, which has mass 1.0, and the fundamental partons are
massless.  One would therefore expect the lowest bound state in the 
spectrum to have a mass of order 1.0.  In ${\cal N}=1$ SYM-CS
theory~\cite{Hiller:2002cu,Hiller:2002pj}, we previously found
anomalously light states at strong coupling, but all of these states were at
or close to threshold. Here the lowest mass state is anomalously light, in
fact nearly massless, and therefore well below threshold.

\begin{figure}
\centerline{
\includegraphics[width=7.5cm]{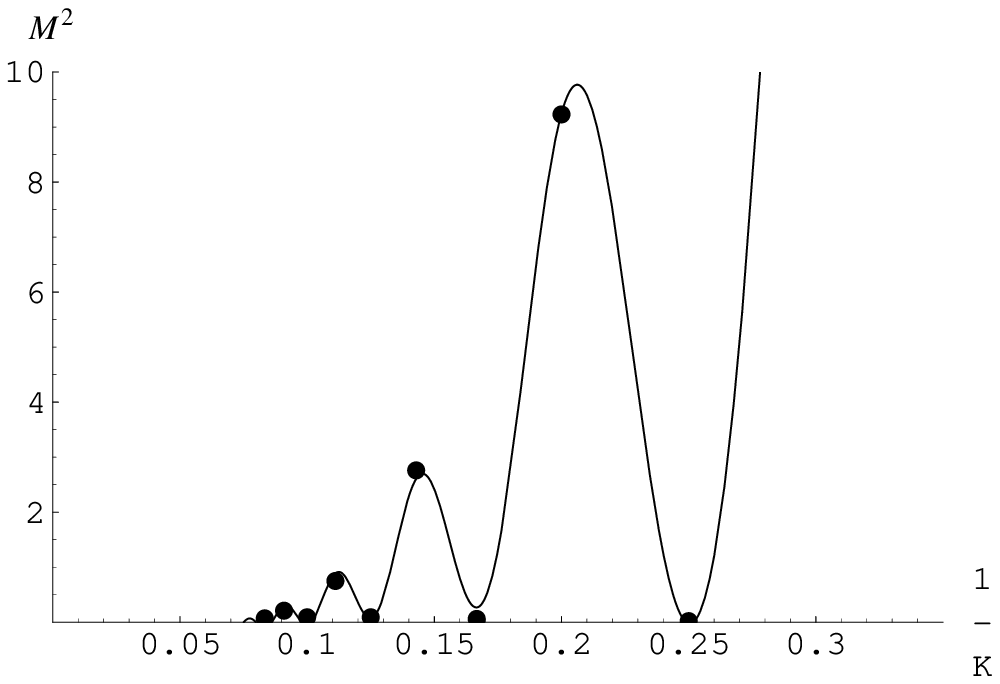}}
\caption{The mass squared of the lowest mass
oscillatory state in units of
$g^2N_c/\pi$, as a function of $1/K$
for $10\kappa^2=g^2 N_c/\pi$ in the $Z_2$-even sector.}
\label{fig:068}
\end{figure}

At yet stronger coupling, $g^2 N_c/\pi=10\kappa^2$, we find that oscillations
are significantly stronger, as seen in Fig.~\ref{fig:068}, and at $K=12$ the
curve is not totally converged. The mass squared at $K=12$ is 0.068, but
without complete convergence we are not able to say anything precise about 
the variation of the bound-state mass as a function of the coupling from
$g^2 N_c/\pi=\kappa^2$ to $g^2N_c/\pi=10\kappa^2$.

%%%%%%%%%%%%%%%%%%%%%%%%%%%%%%%%%%%%%%%%%%%%%%%%%%%%%%%%%%
\subsection{Structure functions}
%%%%%%%%%%%%%%%%%%%%%%%%%%%%%%%%%%%%%%%%%%%%%%%%%%%%%%%%%%

\begin{figure}
\begin{tabular}{cc}
\includegraphics[width=7.5cm]{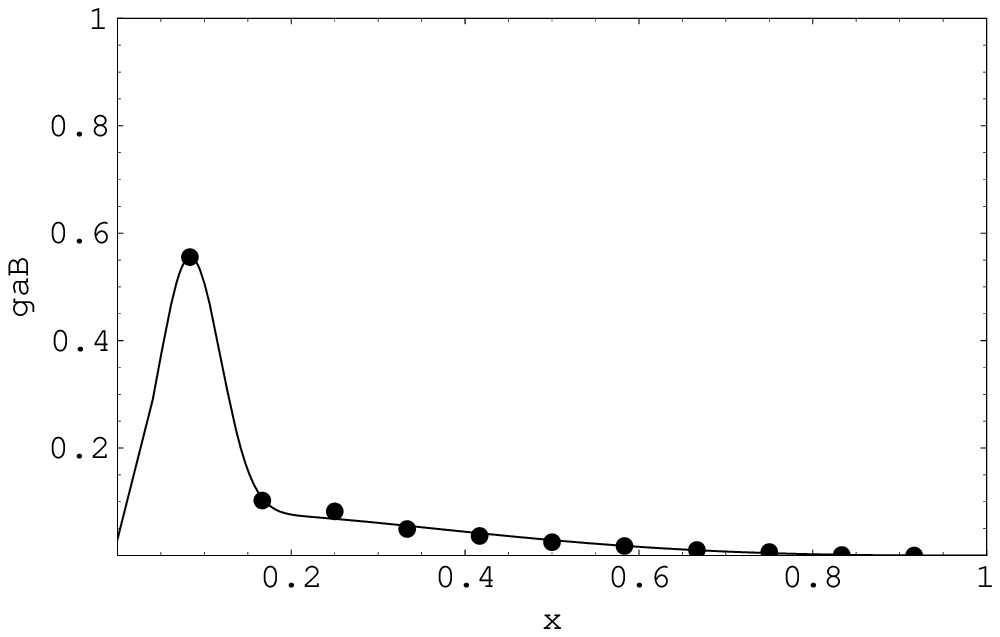}&
\includegraphics[width=7.5cm]{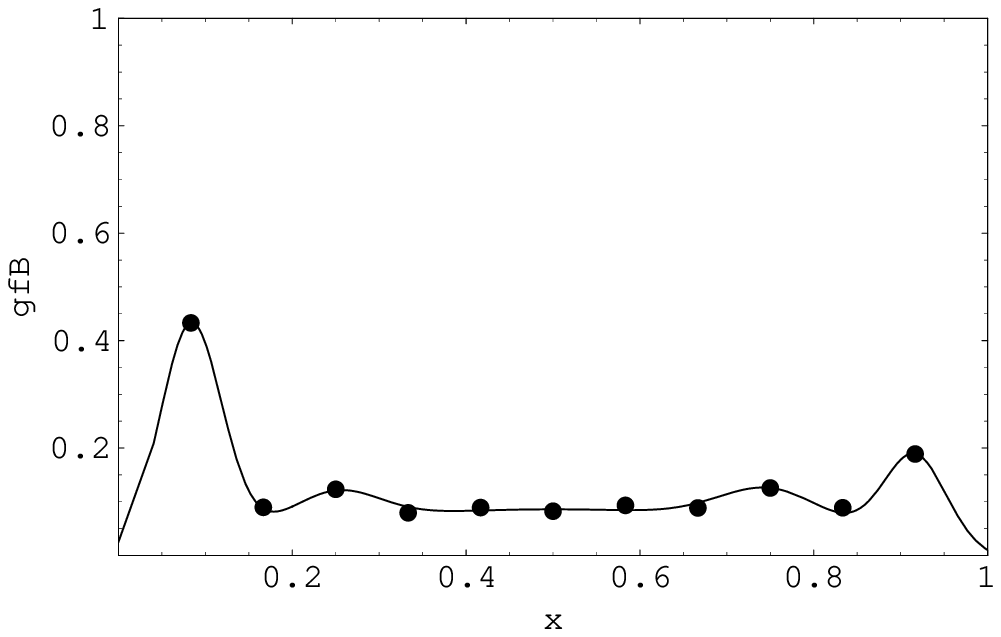}\\
(a) & (b)
\end{tabular}
\caption{Structure functions of the  bound state $M^2=0.025$ at
resolution $K=12$, with $Z_2$ even, for (a) adjoint bosons $aB$ and
(b) fundamental bosons $fB$, with the CS coupling fixed at
$\kappa=g\sqrt{N_c/\pi}$.  The solid curve is a fit to
the computed points.}
\label{fig:025}
\end{figure}

The two largest structure functions for the lowest mass state are shown in
Fig.~\ref{fig:025}.  Even though this is an exactly supersymmetric
theory, nearly all the adjoint partons are gluons.  About
2/3 of the wave function of this state is composed of two fundamental
bosons and an adjoint parton, and about 1/3 of the wave function is made
of two fundamental fermions. Within the context of the standard model,
this state is  primarily a bound state of two squarks and a gluon.
We see that both of these distributions are peaked at small $x$. This
reflects strong binding of the fundamental partons, allowing them to be
widely separated in momentum, combined with only a small contribution
to the momentum from the adjoint boson.

\begin{figure}
\begin{tabular}{cc}
\includegraphics[width=7.5cm]{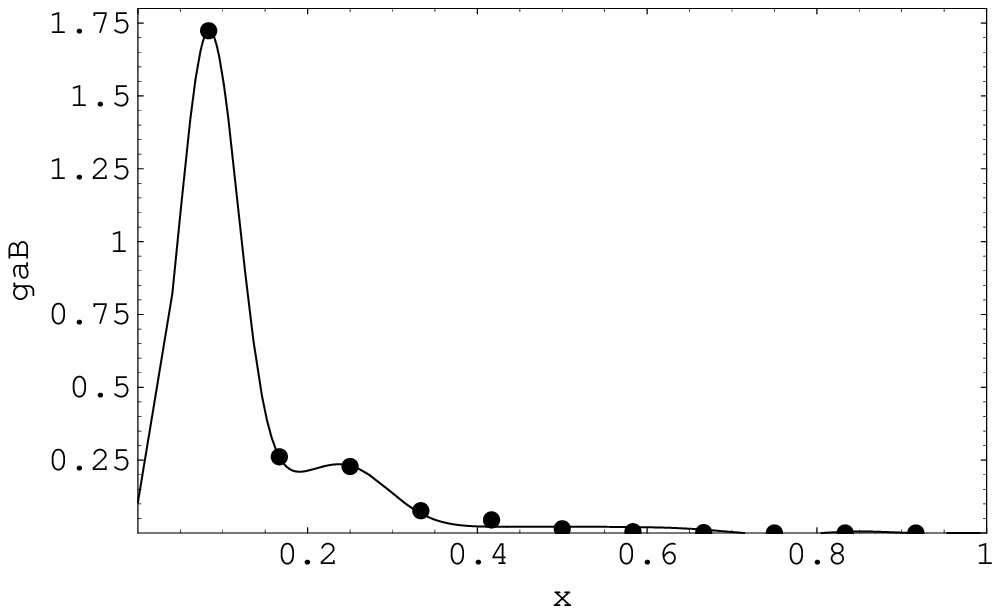}&
\includegraphics[width=7.5cm]{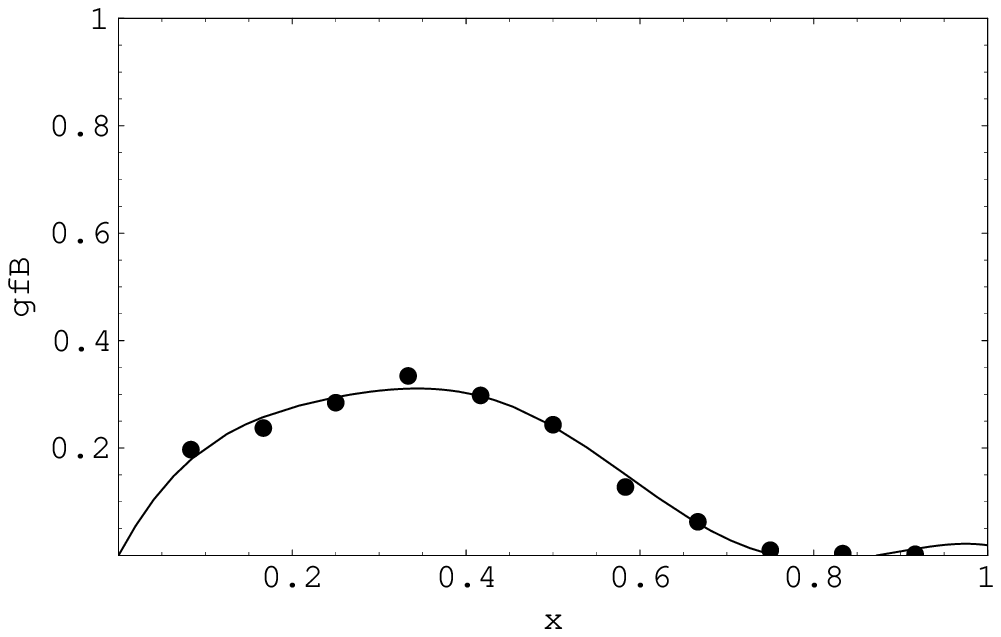}\\
(a) & (b)
\end{tabular}
\caption{Same as Fig.~\ref{fig:025}, but for $M^2=0.29$.
%Structure functions of the  bound state $M^2=0.29$ at
%resolution $K=12$, with $Z_2$ even, for (a) adjoint bosons $aB$ and
%(b) fundamental bosons $fB$, with the CS coupling fixed at
%$\kappa=g\sqrt{N_c/\pi}$.  The solid curve is a fit to
%the computed points.
}
\label{fig:29}
\end{figure}

The average number of partons in the next lowest state is about 4.5\@.
For this state, 90\% is composed of gluons and squarks, and the structure
functions are shown in Fig.~\ref{fig:29}. The squarks have a significantly
wider distribution than in the lowest state.  The gluons remain very
sharply peaked at small $x$.


\begin{figure}
\begin{tabular}{cc}
\includegraphics[width=7.5cm]{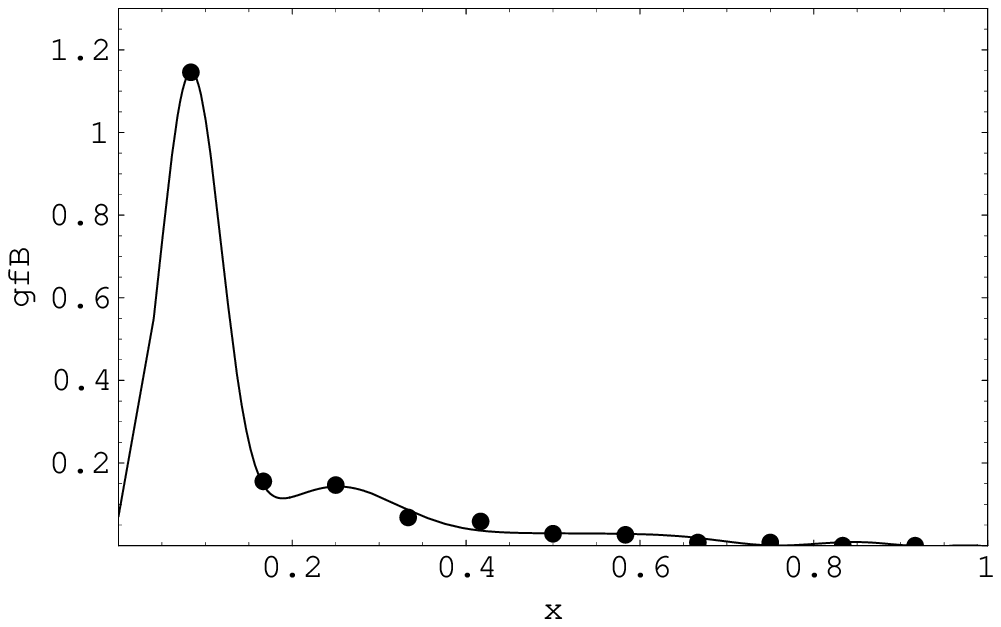}&
\includegraphics[width=7.5cm]{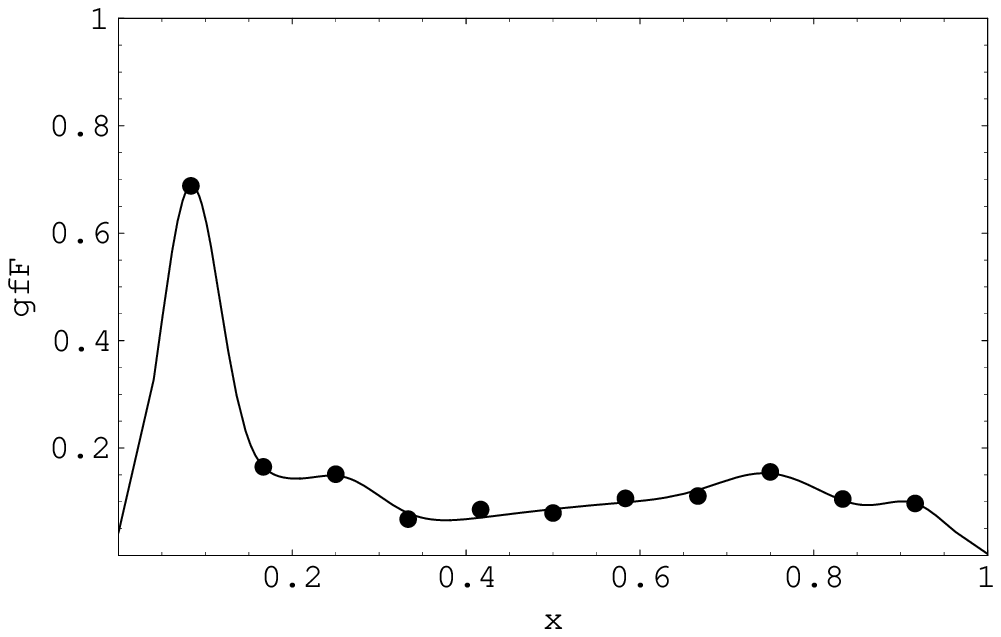}\\
(a) & (b)
\end{tabular}
\caption{Structure functions of the  bound state $M^2=0.135$ at
resolution $K=12$, with $Z_2$ odd, for (a) fundamental bosons $fB$ and
(b) fundamental fermions $fF$, with the CS coupling fixed at
$\kappa=g\sqrt{N_c/\pi}$.  The solid curve is a fit to
the computed points.}
\label{fig:135}
\end{figure}

For the lowest state in the low-mass band with $Z_2$ odd, the average number
of partons is slightly less that 4\@. Again the partons in the adjoint
representation are almost entirely bosons. However, the partons in the
fundamental representation are almost entirely quarks, so this is the
lightest standard-model meson. This is a deeply bound state, since the
gluons have mass 1 and on average there are close to two gluons per
bound state. The structure functions for the gluon and quarks are shown in
Fig.~\ref{fig:135}. As we have seen in all the states that we have
considered, the gluons are strongly peaked at $x$ near zero. The quarks have 
a strong peak at small $x$ but have a significant distribution at other 
values of $x$.

%%%%%%%%%%%%%%%%%%%%%%%%%%%%%%%%%%%%%%%%%%%%%%%%%%%%%%%%%%%%%%%%%
\subsection{Spectra and structure functions in the upper band}
%%%%%%%%%%%%%%%%%%%%%%%%%%%%%%%%%%%%%%%%%%%%%%%%%%%%%%%%%%%%%%%%%

\begin{figure}
\begin{tabular}{cc}
\includegraphics[width=7.5cm]{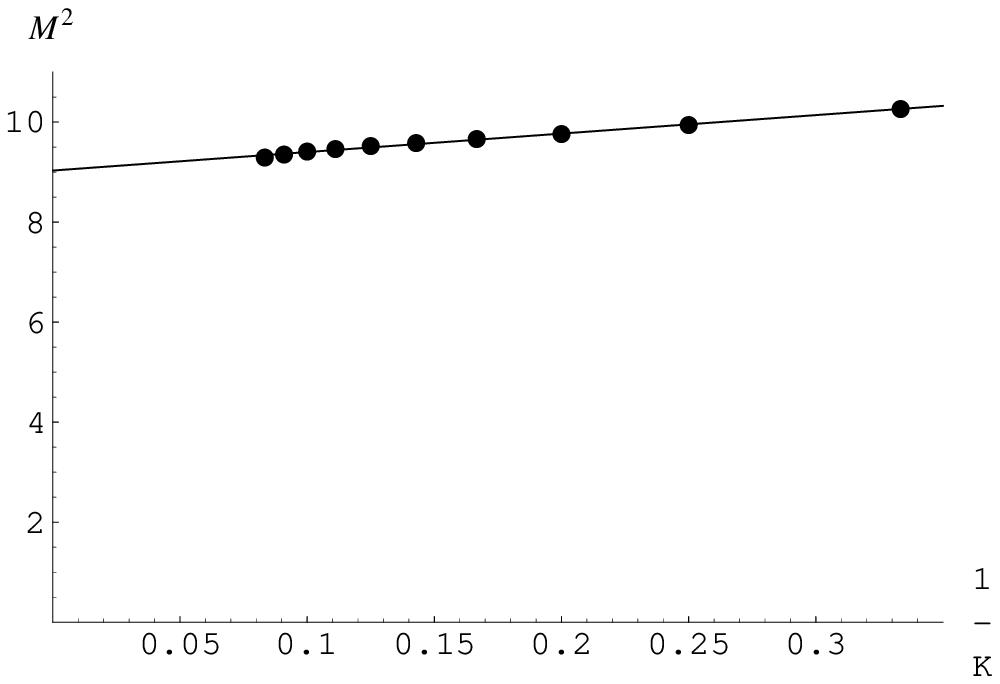}&
\includegraphics[width=7.5cm]{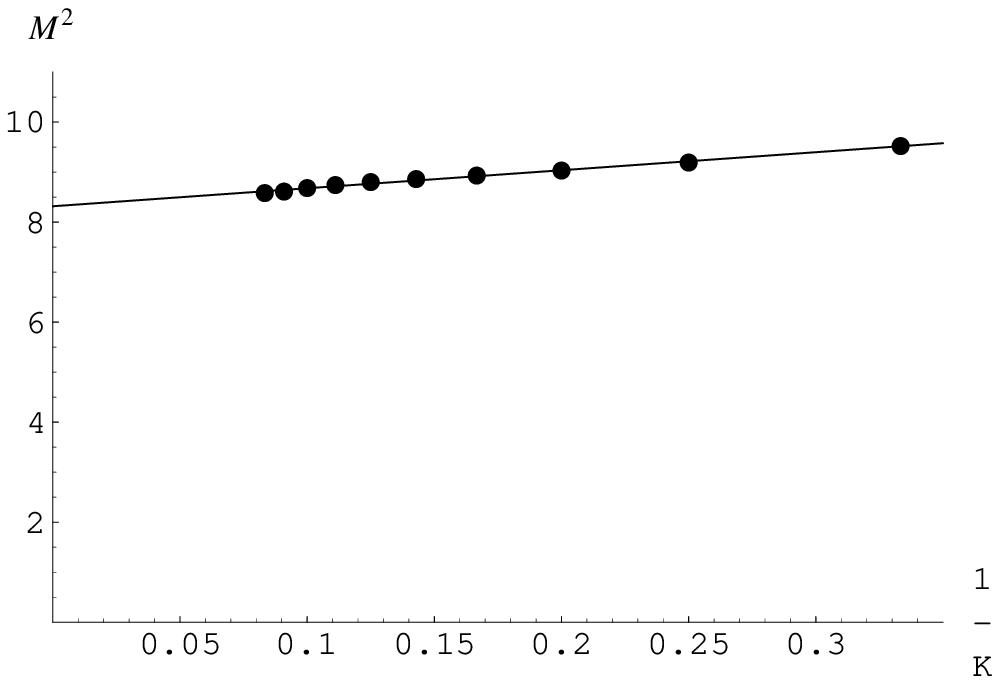}\\
(a) & (b)\\
\end{tabular}
\caption{The mass squared in units of
$g^2N_c/\pi$, as a function of $1/K$
for $\kappa=g\sqrt{N_c/\pi}$ in the $Z_2$-even sector,
of (a) the second lowest mass state of the upper band, with $M^2_\infty=9.03$,
and (b) the  lowest mass state of the upper band, with $M^2_\infty=8.31$. The
solid curve is a fit to the computed points.}
\label{fig:upZ+}
\end{figure}

\begin{figure}
\begin{tabular}{cc}
\includegraphics[width=7.5cm]{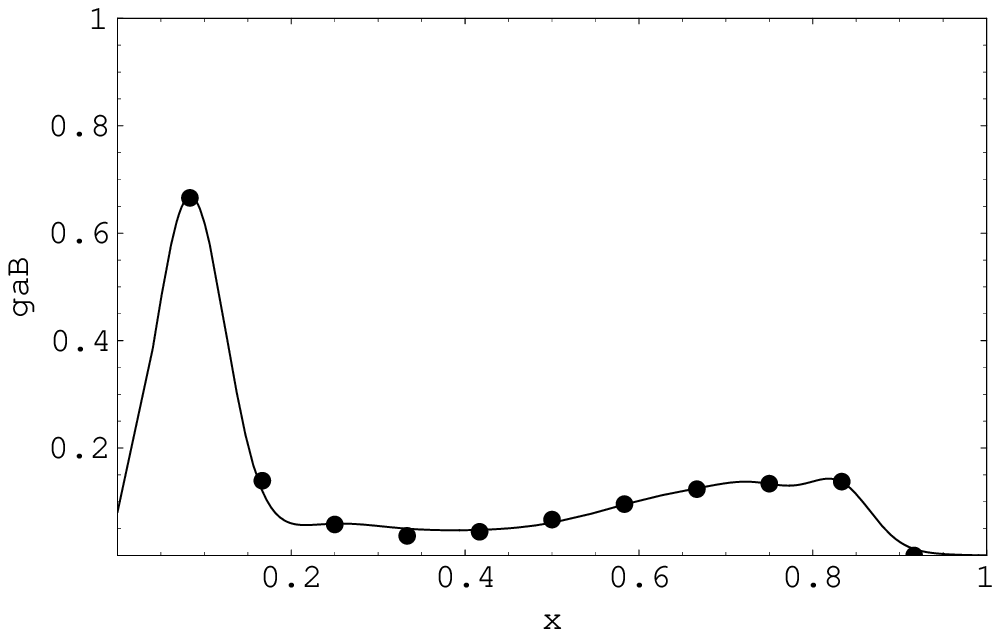}&
\includegraphics[width=7.5cm]{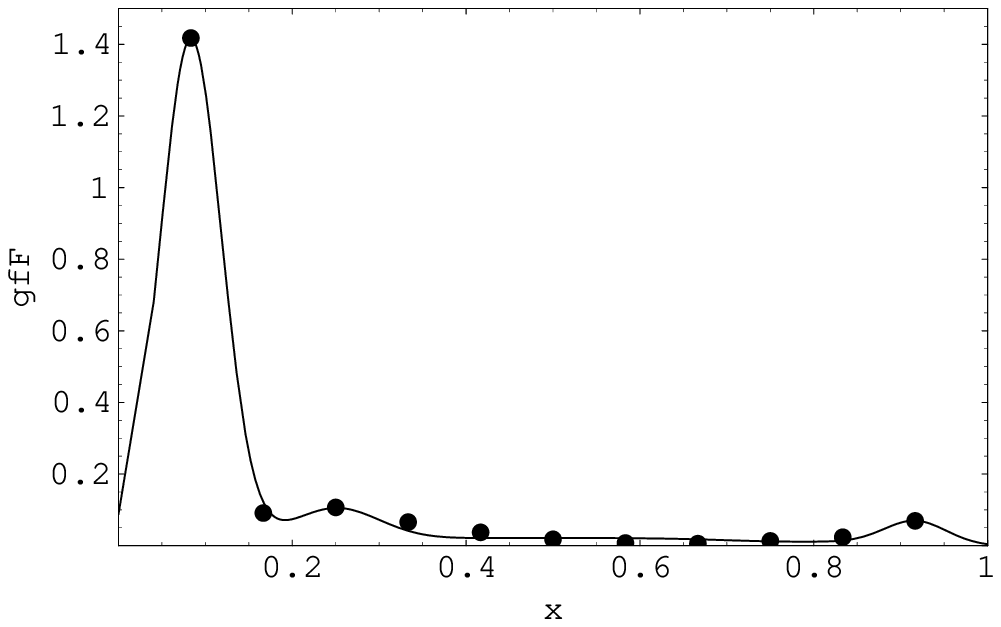}\\
(a) & (b)
\end{tabular}
\caption{Structure functions of the bound state $M^2_\infty=8.31$,
with $Z_2$ even, for (a) adjoint bosons $aB$ and
(b) fundamental fermions $fF$,
with the CS coupling fixed at $\kappa=g\sqrt{N_c/\pi}$.  
The solid curve is a fit to the computed points.}
\label{fig:8.53}
\end{figure}

\begin{figure}
\begin{tabular}{cc}
\includegraphics[width=7.5cm]{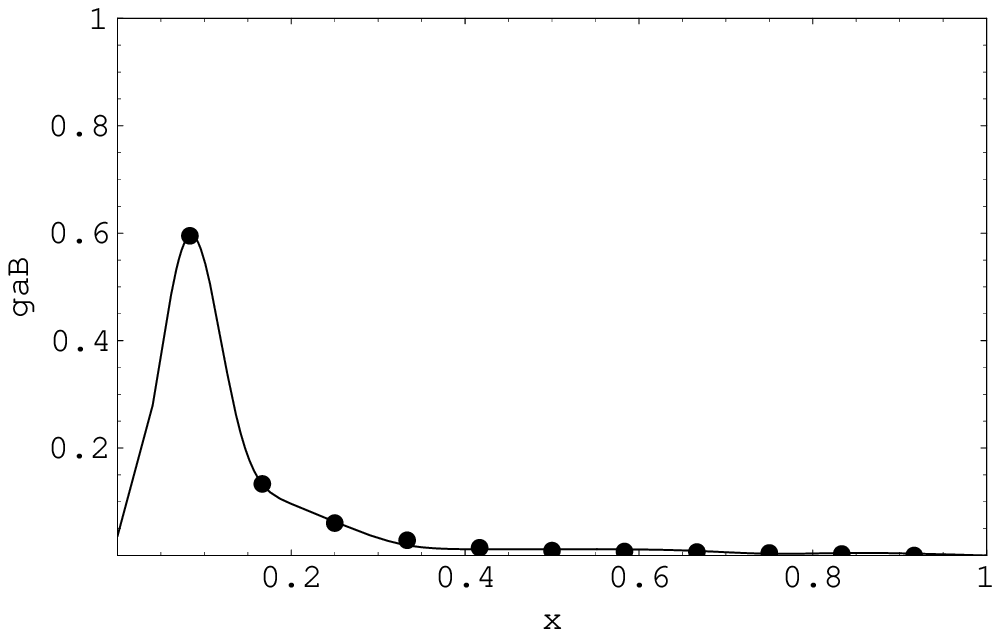}&
\includegraphics[width=7.5cm]{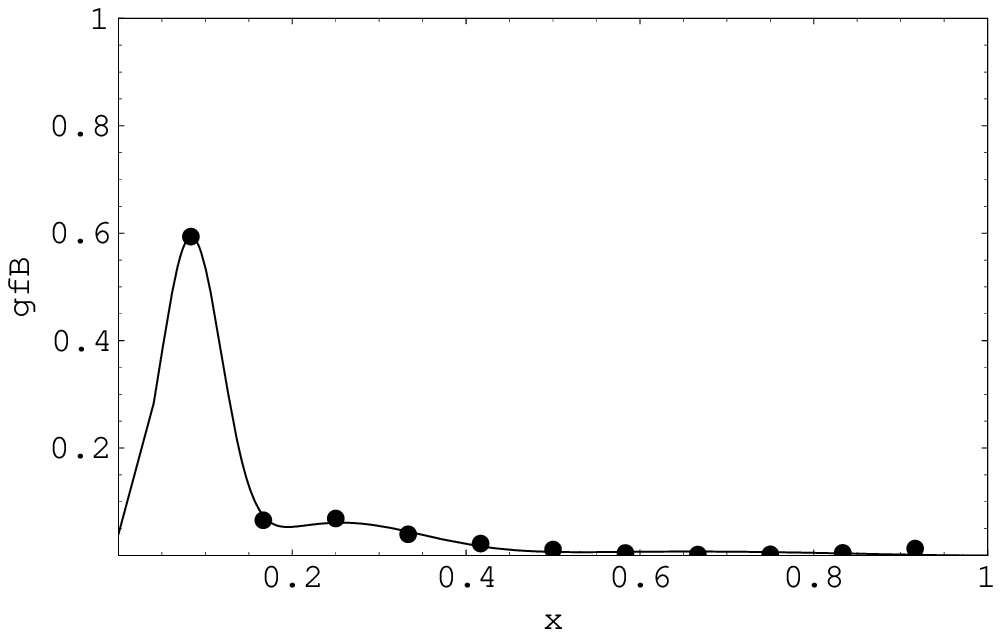}\\
(a) & (b)\\
\includegraphics[width=7.5cm]{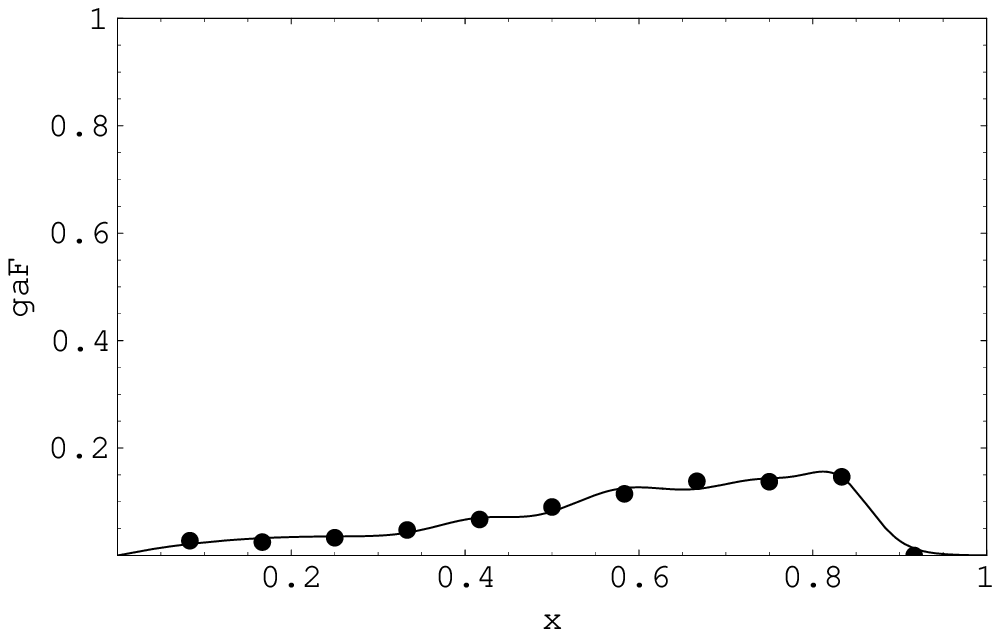}&
\includegraphics[width=7.5cm]{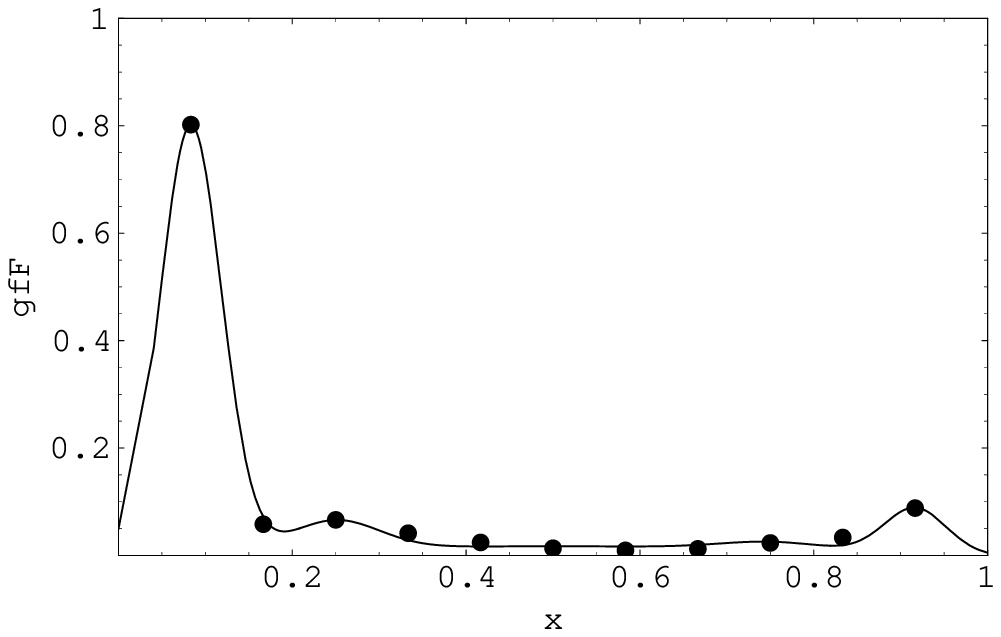}\\
(c) & (d)
\end{tabular}
\caption{
Structure functions of the bound state $M^2_\infty=9.03$ for
(a) adjoint bosons $aB$, (b) fundamental bosons $fB$,
(c) adjoint fermions $aF$, and (d) fundamental fermions $fF$,
with the CS coupling fixed at $\kappa=g\sqrt{N_c/\pi}$ and $Z_2$ even.
The solid curve is a fit to the computed points.
}
\label{fig:9.28}
\end{figure}

While the lower mass band appears to have a very unusual behavior as a
function of the resolution, the bound states in the upper band behave very
similarly to the bound states found in most SDLCQ calculations. The
convergence is excellent, starting from the lowest resolution, and the 
data is fit very well as a function of $1/K$ by a line with a
small slope. In Fig.~\ref{fig:upZ+} we show that the lowest two bound 
states in the upper band
converge to masses $M^2_\infty=8.31$ and $M^2_\infty=9.03$ at infinite
resolution. We carry the calculation out to resolution $K=13$, but the
structure functions are calculated at $K=12$. From the structure
functions of the state at $M^2_\infty=8.31$, we find that this state is 85\%
quarks and gluons, so it is very much like a standard-model meson. The
average number of partons is 3.98, so for the most part the bound state
contains two gluons. It is very interesting that this bound
state is very much like a QCD meson and is the solution of a supersymmetric
field theory. The structure functions are shown in Fig.~\ref{fig:8.53}.
While the gluons are peaked at small $x$, they have a much larger spread than 
we saw in the lower band. The fundamental fermions are very sharply peaked at
small $x$. The structure functions of the state at $M^2_\infty=9.03$ are
interesting because the state has nearly equal mixtures of all four
constituents. The structure functions are
shown in Fig.~\ref{fig:9.28}. The shapes are different, particularly for
the distribution of adjoint fermions, which are spread over the region
of larger $x$.

\begin{figure}
\begin{tabular}{cc}
\includegraphics[width=7.5cm]{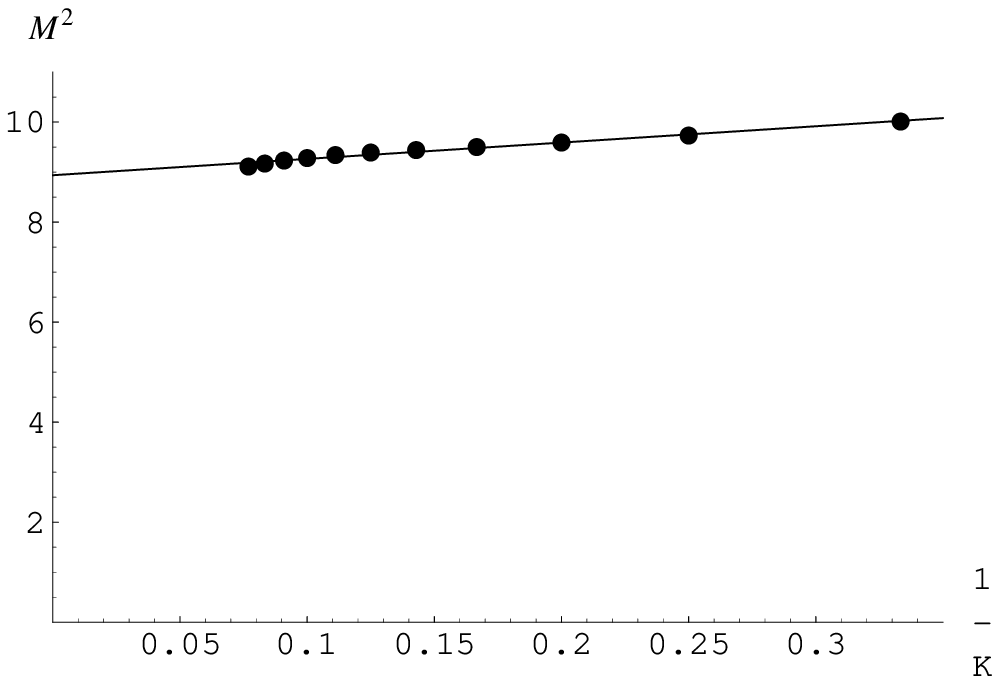}&
\includegraphics[width=7.5cm]{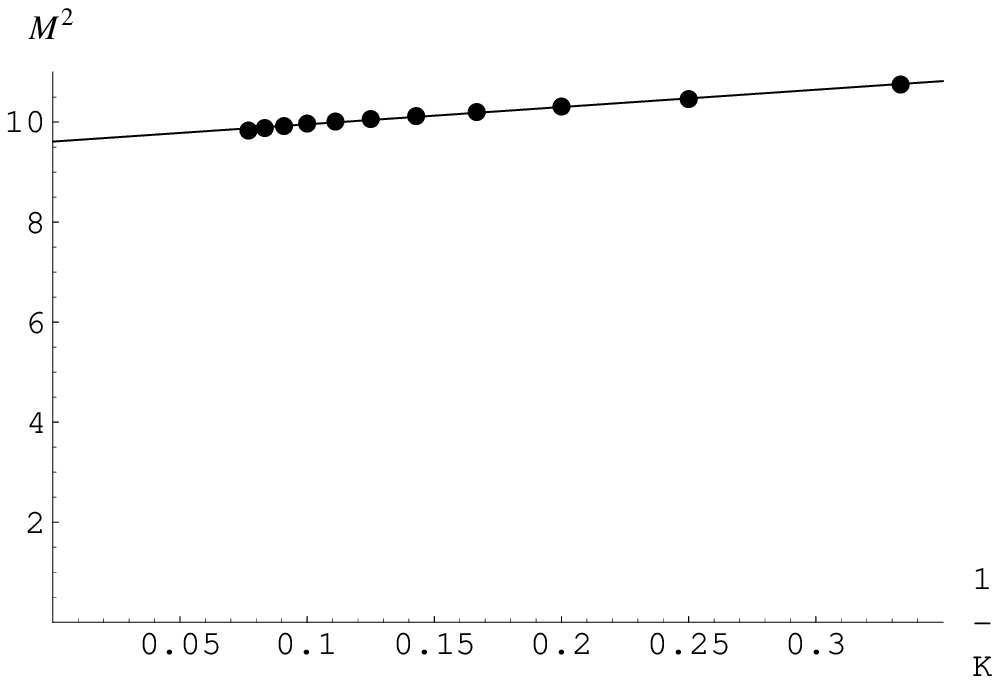}\\
(a) & (b)\\
\end{tabular}
\caption{The mass squared in units of
$g^2N_c/\pi$, as a function of $1/K$
for $\kappa=g\sqrt{N_c/\pi}$ in the $Z_2$-odd sector,
for (a) the  lowest mass state of the upper band, with $M^2_\infty=8.93$,
and (b) the second lowest mass state of the upper band, with
$M^2_\infty=9.6$. The solid curve is a fit to the computed points.
}
\label{fig:upZ-}
\end{figure}

\begin{figure}
\begin{tabular}{cc}
\includegraphics[width=7.5cm]{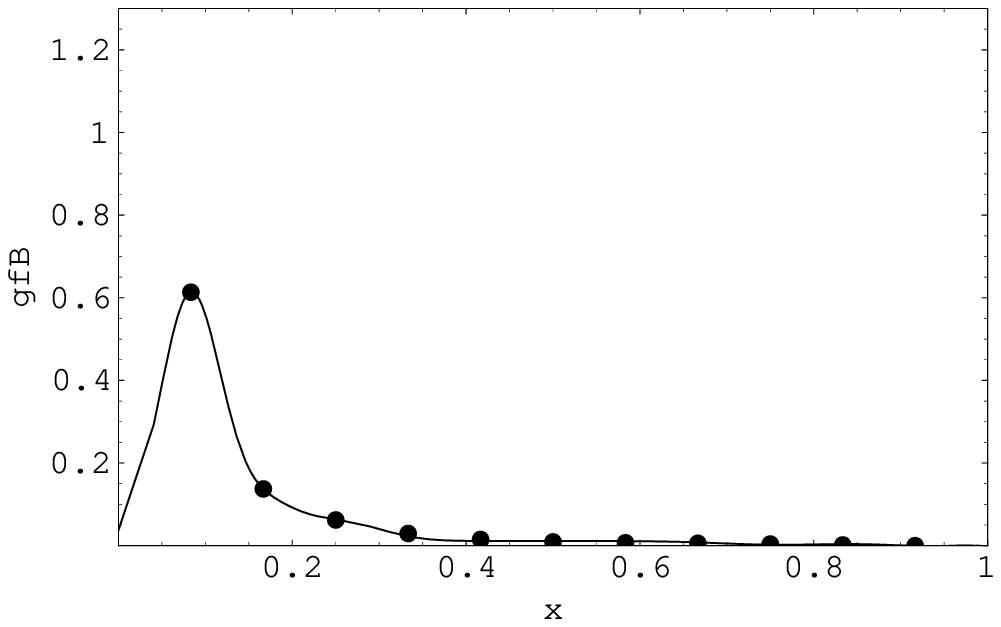}&
\includegraphics[width=7.5cm]{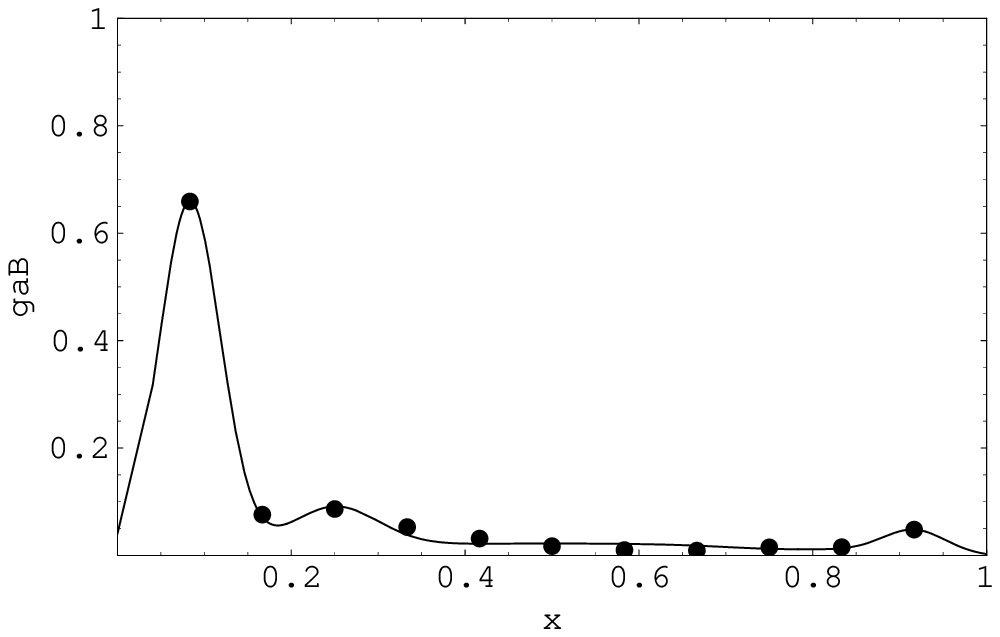}\\
(a) & (b)\\
\includegraphics[width=7.5cm]{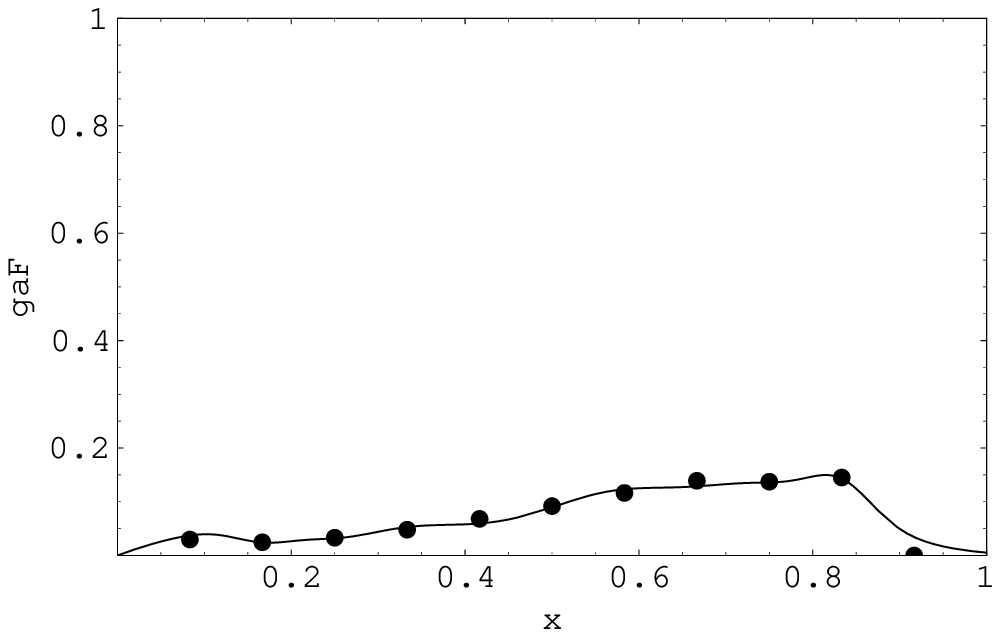}&
\includegraphics[width=7.5cm]{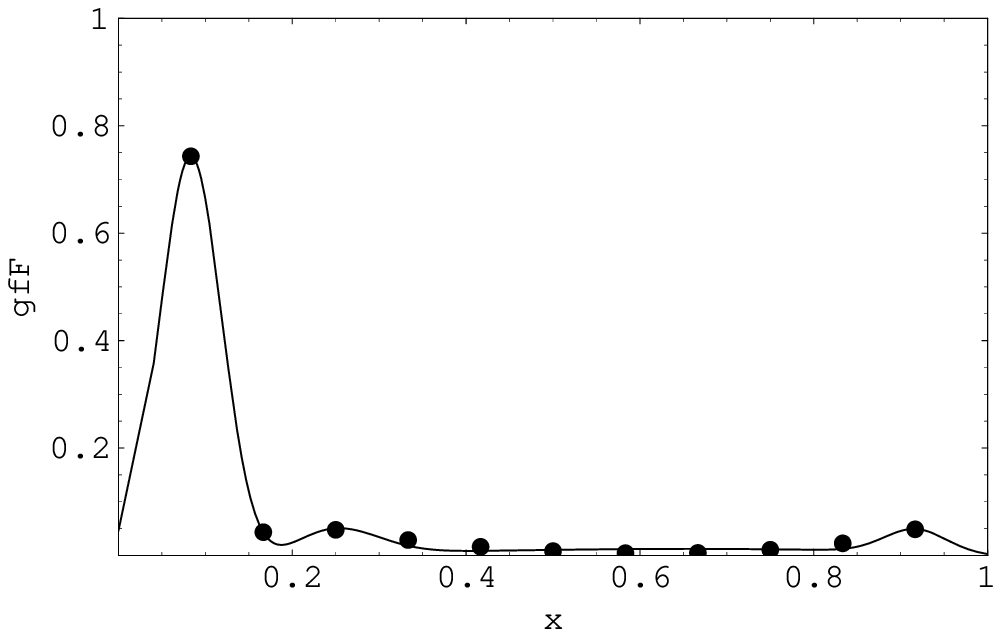}\\
(c) & (d)
\end{tabular}
\caption{Same as Fig.~\ref{fig:9.28}, but for $M^2_\infty=8.93$ and $Z_2$ odd.
%Structure functions of the bound state $M^2_\infty=8.93$ for
%(a) adjoint bosons $aB$, (b) fundamental bosons $fB$,
%(c) adjoint fermions $aF$, and (d) fundamental fermions $fF$,
%with the CS coupling fixed at $\kappa=g\sqrt{N_c/\pi}$ and $Z_2$ odd.
%The solid curve is a fit to the computed points.
}
\label{fig:9.17}
\end{figure}

In Fig.~\ref{fig:upZ-} we see that in the $Z_2$-odd sector we
again have an excellent linear fit.  The two lowest states have
masses of $M^2_\infty=8.93$ and $M^2_\infty=9.60$. The structure 
functions for this sector are similar to those for the $Z_2$-even
states. The structure functions for the state with
$M^2_\infty=8.93$  are shown in Fig.~\ref{fig:9.17} and are
very similar to the structure functions for the state with
$M^2_\infty=9.03$ in the
$Z_2$-even sector. 

%%%%%%%%%%%%%%%%%%%%%%%%%%%%%%%%%%%%%%%%%%%%%%%%%%%%%%%%%
\section{Discussion.}
\label{secdis}
%%%%%%%%%%%%%%%%%%%%%%%%%%%%%%%%%%%%%%%%%%%%%%%%%%%%%%%%%

In this paper we studied ${\cal N}=1$ SYM-CS theory with
fundamental matter in 1+1 dimensions. The CS term was included to give
masses to the adjoint partons.  The calculations were performed at large
$N_c$ in the framework of SDLCQ; namely, we compactified the light-like
coordinate $x^-$ on a finite circle and calculated the Hamiltonian as the
square of a supercharge $Q^-$, which we then diagonalized numerically.
We found that the spectrum of this theory has two bands, a lower mass
band and an upper mass band. With a CS term present, we found that these bands
separate for $g^2N_c/\pi=\kappa^2$, and we can easily study them separately.

For very massive adjoint partons, the lower mass band becomes a set of massless
bound states composed of only fundamental partons. When the mass for the
adjoint partons is reduced, the massless states become the states of the
lower mass band. The states in the low-mass band are unusual in that their
convergence is oscillatory.  We show that the upper and lower bounds of the
low-mass band have the same shape as the DLCQ approximation to the
two-particle free spectrum, which has an oscillatory behavior for its lower
bound and a growing upper bound.  We argue therefore that the oscillatory
behavior is a numerical remnant of the free two-particle spectrum. In
previous work~\cite{Hiller:2003qe} we found that at $g^2 N_c/\pi=\kappa^2$ 
the lowest mass state of this theory was anomalously low, and in fact close
to zero, while threshold for this state is at $M^2=1$. On average this state 
has one massive gluon and two squarks. This is the fifth supersymmetric 
theory where we have found anomalously light states. In SYM in 1+1~\cite{alp2}
and 2+1~\cite{Antonuccio:1999zu} dimensions, there are massless BPS
states. In SYM-CS theory in 1+1 dimensions we saw that at strong coupling
the lightest states are approximately BPS states whose masses are
independent of the YM coupling~\cite{Hiller:2002cu}.  In SYM-CS theories 
in 2+1 dimensions at strong coupling, there is again an anomalously light 
bound state~\cite{Hiller:2002pj}.  The next lightest state in this band has
about four partons. We carry the calculation to resolution $K=13$, so in
principle these states could contain 13 partons. In addition,
almost all of the partons in the adjoint representation are gluons rather
than gluinos.  Also, the content of the particles in the fundamental
representation is rather pure, either squarks or quarks. Thus some of 
the boson bound states are rather pure combinations of quarks and glue, 
in spite of the fact that they are bound states of an exactly 
supersymmetric theory.

The bound states in the upper mass band are rather conventional SDLCQ bound
states.  They converge very quickly and have the conventional linear
behavior in $1/K$.  Similar to the lower mass band, most of the partons 
in the adjoint representation are gluons. In addition, the states are 
again rather pure quark or squark states. When we take the adjoint parton 
mass large, these are the states that have at least one adjoint parton.

We calculate the structure functions of the states in both bands for all
four species of partons, and we find that most of the distributions are 
strongly peaked at $x$ near zero. This is an indication that for $g^2
N_c/\pi=\kappa^2$ 
these are strongly coupled bound states. In a weakly coupled bound state one
would, on average, expect the partons to share the momentum fraction
equally, and therefore the structure function would be peaked at 1/3 for a
state with three partons, for example. Since several of the states we
consider are almost pure quark-gluon bound states, they might not be 
affected by supersymmetry breaking, which would give large mass to the 
squarks and gluinos.  It would be interesting to investigate in more 
detail the properties of these states and compare them with QCD.

There remains a considerable amount of work to be done on SYM-CS theories
with fundamental matter. The most straightforward extension of the
present work is to consider calculations in 2+1
dimensions~\cite{Antonuccio:1999zu,Haney:2000tk,hpt2001}.  The
${\cal N}=1$ theory in 2+1 dimensions is easily within our reach. Beyond
that the ${\cal N}=2$ theory~\cite{Antonuccio:1998mq} in 2+1 dimensions, 
which is the dimensional reduction of the ${\cal N}=1$ theory in 3+1 
dimensions, will be very interesting. For finite $N_c$ we could also look 
at the mixing of closed-loop states and finite string-like meson states. 
For $\kappa=0$ the bound states of all these theories provide an interesting 
field-theoretic model for strings.

%%%%%%%%%%%%%%%%%%%%%%%%%%%%%%%%%%%%%%%%%%%%%%%%%%%%%%%%%%
\acknowledgments
This work was supported in part by the U.S. Department of Energy.
%%%%%%%%%%%%%%%%%%%%%%%%%%%%%%%%%%%%%%%%%%%%%%%%%%%%%%%%%%

\end{document}